\documentclass{emulateapj}
\shorttitle{BIPOLAR SHOCK BREAKOUT}
\shortauthors{SUZUKI,\ MAEDA,\ \&SHIGEYAMA}
\begin{document}
\title{2D RADIATION-HYDRODYNAMIC SIMULATIONS OF SUPERNOVA SHOCK BREAKOUT IN BIPOLAR EXPLOSIONS OF A BLUE SUPERGIANT PROGENITOR}
\author{AKIHIRO SUZUKI\altaffilmark{1}, KEIICHI MAEDA\altaffilmark{1,2}, and TOSHIKAZU SHIGEYAMA\altaffilmark{3}}
\altaffiltext{1}{Department of Astronomy, Kyoto University, Kitashirakawa-Oiwake-cho, Sakyo-ku, Kyoto, 606-8502, Japan}
\altaffiltext{2}{Kavli Institute for the Physics and Mathematics of the Universe (WPI),
The University of Tokyo, 5-1-5 Kashiwanoha, Kashiwa, Chiba 277-8583, Japan}
\altaffiltext{3}{Research Center for the Early Universe, School of Science, University of Tokyo, Bunkyo-ku, Tokyo, 113-0033, Japan.}
\begin{abstract}
A two-dimensional special relativistic radiation-hydrodynamics code is developed and applied to numerical simulations of supernova shock breakout in bipolar explosions of a blue supergiant. 
Our calculations successfully simulate the dynamical evolution of a blast wave in the star and its emergence from the surface.  
Results of the model with spherical energy deposition show a good agreement with previous simulations. 
Furthermore, we calculate several models with bipolar energy deposition and compare their results with the spherically symmetric model. 
The bolometric light curves of the shock breakout emission are calculated by a ray-tracing method. 
Our radiation-hydrodynamic models indicate that the early part of the shock breakout emission can be used to probe the geometry of the blast wave produced as a result of the gravitational collapse of the iron core.
\end{abstract}

%%%%%%%%%%%%%%%%%%%%%%%
%%%    INTRODUCTION
%%%%%%%%%%%%%%%%%%%%%%%
\section{INTRODUCTION\label{intro}}
The first electromagnetic signal from an explosion of a massive star at the final evolutionary state, i.e., a core-collapse supernova (CCSN) explosion, comes from gas heated by a strong shock wave propagating in the atmosphere of the star. 
CCSNe are initiated by the gravitational collapse of the iron core having grown during their evolutions. 
A fraction of the gravitational energy liberated by the collapsing core is deposited at the stellar mantle stratified on the core and then a strong shock wave forms as a result of the energy deposition, sweeps the stellar mantle, and converts the shock kinetic energy into the internal energy of the shocked gas. 
When the shock front is located at a sufficiently deep layer in the star, photons created by the shocked gas cannot escape from the shocked region, owing to the large optical depth from the shock front to the stellar surface. 
However, when the shock wave finally reaches the stellar surface, photons having been trapped in the shocked region start leaking out to the interstellar space. 
This phenomenon can be observed as an intense outburst of UV or X-ray photons and called supernova shock breakout. 

Although the association of bright and high-energy emission with the birth of a CCSN has been theoretically pointed out by several earlier studies in the 1970s \citep{1974ApJ...187..333C,1978ApJ...223L.109K,1978ApJ...225L.133F}, detection of shock breakout emission has been a big challenge for a long time because of the difficulty of identifying rapidly fading transients as soon as they appear in the sky. 
We do not know where and when a new CCSN appears in the sky in advance, which means that strategic blind survey projects covering a wide region of the whole sky at high cadence are desired. 
Despite the difficulty, the development of modern telescopes and all-sky survey missions as well as the sophistication of observational techniques have eventually made it possible to detect the early electromagnetic signal from CCSNe. 
Since the serendipitous discovery of an outburst of X-ray photons spatially and temporarily coincident with the birth of a SNIb 2008D \citep{2008Natur.453..469S}, an increasing number of possible associations of bright UV flashes with CCSNe have been reported \citep{2008Sci...321..223S,2008ApJ...683L.131G,2010ApJ...724.1396O,2015ApJ...804...28G}. 
Furthermore, the high brightness of the shock breakout emission in optical and UV bands makes it a promising tracer for star-forming activity in the high-z universe \citep{2011ApJS..193...20T}.

From a theoretical point of view, a lot of works have been done since the pioneering works, aiming at clarifying the behavior of the shock wave in the stellar envelope during the shock emergence. 
\cite{1999ApJ...510..379M} investigated how the blast wave produced as a result of the explosion propagates in the stellar interior and finally expels the envelope in both analytical and numerical ways based on a self-similar solution describing the shock emergence in a planar atmosphere \citep{S60}. 
Supernova shock breakout in a relativistic regime recently receivs an increasing attention owing to their potential to produce high-energy emission. 
\cite{2001ApJ...551..946T} performed an analysis similar to {\cite{1999ApJ...510..379M} in a relativistic regime. 
\cite{2005ApJ...627..310N} found a self-similar solution corresponding to the ultra-relativistic extension of Sakurai's solution.

Furthermore, several studies analytically derived formulae for shock breakout light curves from optical to UV \citep[e.g.,][]{1992ApJ...394..599C,2007ApJ...667..351W,2008ApJ...683L.135C,2010ApJ...708..598P,2010ApJ...725..904N,2011ApJ...728...63R,2012ApJ...747...88N,2012ApJ...747..147K}, which can be compared with observed light curves. 

Around a strong shock propagating in a stellar envelope, photons created in the downstream diffuse into the upstream, leading to the deceleration of the downstream flow, i.e., the shock is mediated by radiation \citep{1976ApJS...32..233W}. 
Therefore, it is important to investigate the structure of a radiation-mediated shock in understanding high-energy radiation produced in supernova shock breakout. 
Recently, several studies were carried out to unveil the structure of radiation-mediated shocks in the context of supernova shock breakout \citep{2010ApJ...716..781K,2010ApJ...725...63B} and claim that the shocked gas immediately after the shock passage can be very hot due to inefficient coupling between radiation and gas and produce X-ray and gamma-ray photons. 
Repeated Compton scatterings around the shock front may also play a role in producing non-thermal photons when the shock velocity is mildly relativistic or larger \citep{2007ApJ...664.1026W,2010ApJ...719..881S}. 
Recently, \cite{2013ApJ...777..113O} combined radiative transfer calculations based on a Monte Carlo technique with the self-similar solution describing ultra-relativistic shock breakout \citep{2005ApJ...627..310N} and demonstrated that MeV photons can be produced when the shock velocity is highly relativistic. 
These proposed mechanisms may be responsible for high-energy radiation observed from some energetic SNe associated with an X-ray flash, such as, SNe 2006aj \citep{2006Natur.442.1008C,2006Natur.442.1011P,2006Natur.442.1014S,2006Natur.442.1018M}, 2008D, and 2010bh \citep{2011MNRAS.411.2792S}.

Numerical modelings of supernova shock breakout are also of great importance to clarify the dynamical evolution of the strong shock before and after the breakout and calculate expected emission from the shocked gas. 
Since the detection of SN 1987A in the Large Magellanic Cloud, several spherically symmetric hydrodynamic simulations including radiative transfer have been done to reproduce observed properties of this particular event \citep{1990ApJ...360..242S,1992ApJ...393..742E}. 
Furthermore, recent sophisticated numerical models including multi-group radiative transfer shed light on the spectral evolution of the shock breakout emission in explosions of a wide variety of progenitor stars from Wolf-Rayet stars to blue and red supergiants \citep{2000ApJ...532.1132B,2009ApJ...705L..10T,2011ApJ...742...36S,2013MNRAS.429.3181T,2013ApJ...774...79S,2014ApJ...796..145S}.

Furthermore, as suggested from recent observations of CCSNe at very early epochs, they might be explosions of stars with extended envelopes or with dense circumstellar media \citep{2013Natur.494...65O,2014Natur.509..471G,2014ApJ...789..104O}. 
The presence of an optically thick medium in the vicinity of the progenitor star can make the shock breakout emission much brighter and longer than the that from the stellar surface \citep[e.g.,][]{1973ApJ...180L..65F,1977ApJS...33..515F,2011ApJ...729L...6C}. 
The early emission from the shock breakout plays crucial roles in investigating the mass-loss episode in the last few hundreds years before the death of massive stars. 
Recently, detailed modelings of light curves and spectra from a shock emerging from an optically thick wind or envelope have been extensively carried out \citep[e.g.,][]{2011MNRAS.414.1715B,2011MNRAS.415..199M,2012ApJ...757..178G,2012ApJ...759..108S,2014ApJ...780...18G,2014ApJ...788..113S,2015A&A...575L..10M}.

Another important factor to possibly modify or prolong shock breakout light curves is multi-dimensional effects. 
The importance of the deviation from spherical symmetry has been recognized and paid attention for a long time from both theoretical and observational viewpoints. 
Observations of CCSNe have revealed that both continuum and line polarization are commonly found in their spectra \citep[see,][for review]{2008ARA&A..46..433W}. 
In the standard theory of the neutrino-driven mechanism for CCSNe, multi-dimensional effects play critical roles in the revival of the stalled shock after the collapse of the iron core \citep[e.g.,][]{2006RPPh...69..971K,2007PhR...442...38J,2012ARNPS..62..407J}.  
Although some authors have considered effects of aspherical explosions on shock breakout emission and pointed out that the explosion geometry can actually affect the emission \citep[e.g.,][]{2010ApJ...717L.154S,2011ApJ...727..104C,2013ApJ...779...60M,2014ApJ...790...71S}, 
these studies are restricted to hydrodynamics without radiative transfer. 
Radiation-hydrodynamic simulations in 2D and 3D could play vital roles in clarifying how an aspherical explosion affects the supernova shock breakout and in establishing multi-dimensional models. 
Thanks to powerful modern parallel computers and the formulation of numerical schemes for radiation-hydrodynamics, it has become possible to run 2D or 3D radiation-hydrodynamic simulations. 
In this paper, we describe a 2D radiation-hydrodynamics code developed by one of the authors and present results of simulations of the shock propagation in a massive star, its emergence from the surface, and the decoupling of radiation and gas after the breakout in both spherical and aspherical explosions. 

This paper is structured as follows. 
In \S 2, we describe details of the numerical code used in this study. 
In \S 3, we present the setups and results of the simulations for supernova shock breakout in a blue supergiant progenitor, which are compared with earlier 1D radiation-hydrodynamic simulations. 
The bolometric light curves of the shock breakout emission are calculated in \S 4. 
Finally, \S 5 concludes this paper.  
In Appendix, results of some test calculations performed to check the validity of the numerical code are shown. 
We use the unit $c=1$ unless otherwise noted. 
The Einstein summation convention is used for repeated indices in equations. 

%%%%%%%%%%%%%%%%%%%%%%%
%%%    FORMULATION
%%%%%%%%%%%%%%%%%%%%%%%
\section{FORMULATION AND NUMERICAL CODE}
In this section, we introduce the basic equations and our method to numerically integrate the equations. 
There have been a lot of works focusing on how to solve equations of hydrodynamics with radiative transfer and the development of numerical codes \citep{1992ApJS...80..819S,2001ApJS..135...95T,2003ApJS..147..197H,2006ApJS..165..188H,2007A&A...464..429G,2007ApJ...667..626K,2008PhRvD..78b4023F,2011ApJS..196...20Z,2013MNRAS.429.3533S,2013ApJ...764..122T,2013ApJ...772..127T} and some of the codes are publicly available. 
Although our numerical scheme is based on these studies and not new, we briefly describe some key aspects of the formulation, the assumptions, and the numerical techniques employed in our code. 
The validity of our code has been checked by solving some test problems, which are reviewed in Appendix of this paper. 

\subsection{Basic Equations}
In the treatment of the equations, we consider two inertial frames, the comoving and laboratory frames. 
Fluid elements are at rest in the former frame, while they are moving in the latter. 
We solve equations of special relativistic radiation-hydrodynamics in the so-called mixed frame \citep[see, e.g.,][]{1984oup..book.....M}, in which the physical variables of the radiation field are defined in the laboratory frame, while the absorption and scattering coefficients are defined in the comoving frame. 
A radiation field is fully described by the intensity, $I_\nu$, which is a function of the time $t$, the coordinates ${\textbf{\textit x}}$ in the physical space, the frequency $\nu$, and the direction vector ${\textbf{\textit l}}$. 
The temporal evolution of the intensity is governed by the transfer equation, which is an advection equation in the phase space. 
However, the direct integration of the transfer equation requires huge computational resources and is not practical. 
Therefore, instead of solving the transfer equation, we consider the temporal evolution of the frequency-integrated radiation energy density $E_\mathrm{r}$ and the radiation flux ${F}^i_\mathrm{r}$ defined as the zeroth and first order angular moments of the intensity integrated over the frequency,
\begin{equation}
E_\mathrm{r}(t,{\textbf{\textit x}})=\int I_\nu d\nu d\Omega,
\label{eq:def_er}
\end{equation}
and
\begin{equation}
F^i_\mathrm{r}(t,{\textbf{\textit x}})=\int I_\nu l^i d\nu d\Omega,
\label{eq:def_fr}
\end{equation}
where the superscript $i$ specifies the component of a vector or tensor. 
Furthermore, the second order moment of the intensity yields the radiation pressure tensor,
\begin{equation}
P^{ij}_\mathrm{r}(t,{\textbf{\textit x}})=\int I_\nu l^il^j d\nu d\Omega.
\end{equation}

The dynamical evolution of radiation and gas should be solved simultaneously. 
The hydrodynamical variables characterizing the gas are the velocity $\beta^i$, the density $\rho$, and the gas pressure $P_\mathrm{g}$, which are functions of the spatial coordinates and the time. 
In this paper, we assume the equation of state for an ideal gas with an adiabatic index $\gamma=5/3$. 
Thus, the specific enthalpy $h$ of the gas is expressed as follows,
\begin{equation}
h=1+\frac{\gamma P_\mathrm{g}}{(\gamma-1)\rho}.
\end{equation}
The gas temperature $T_\mathrm{g}$ is derived for a given set of the density, the pressure, and the chemical composition of the gas from the following equation,
\begin{equation}
P_\mathrm{g}=(\gamma-1)E_\mathrm{g}=
\frac{\rho k_\mathrm{B}T_\mathrm{g}}{\mu m_\mathrm{u}},
\end{equation}
where $k_\mathrm{B}$ and $m_\mathrm{u}$ are the Boltzmann constant and the atomic mass unit and $\mu$ and $E_\mathrm{g}$ denote the mean molecular weight and the internal energy of the gas. 
On the other hand, the radiation temperature $T_\mathrm{r}$ is defined as the temperature obtained from the following relation,
\begin{equation}
E_\mathrm{r}=a_\mathrm{r}T_\mathrm{r}^4,
\end{equation}
where $a_\mathrm{r}$ is the radiation constant. 
These two values should take the same value, $T_\mathrm{r}=T_\mathrm{g}$, when the energy exchange between gas and radiation is balanced. 
 
We denote the frequency-averaged absorption and scattering coefficients in the comoving frame  by $\kappa_0$ and $\sigma_0$. 
Under the mixed-frame formalism \citep[see, e.g.,][]{1984oup..book.....M}, the equations describing the temporal evolutions of hydrodynamical variables, the radiation energy density, the radiation flux are written as follows,
\begin{eqnarray}
\frac{\partial (\rho\Gamma)}{\partial t}+\frac{\partial (\rho\Gamma\beta^j)}{\partial x^j}&=&0,
\label{eq:continuity}
\\
\frac{\partial (\rho h\Gamma^2\beta^i)}{\partial t}+\frac{\partial (\rho h\Gamma^2\beta^i\beta^j+P_\mathrm{g}\delta^{ij})}{\partial x^j}&=&G^i,
\label{eq:momentum}
\\
\frac{\partial (\rho h\Gamma^2-P_\mathrm{g})}{\partial t}+\frac{\partial (\rho h\Gamma^2\beta^i)}{\partial x^j}&=&G^0,
\label{eq:energy}
\\
\frac{\partial E_\mathrm{r}}{\partial t}+\frac{\partial F_\mathrm{r}^i}{\partial x^i}&=&
-G^0,
\label{eq:radiation_energy}
\\
\frac{\partial F_\mathrm{r}^i}{\partial t}+\frac{\partial P_\mathrm{r}^{ij}}{\partial x^j}&=&
-G^i,
\label{eq:radiation_flux}
\end{eqnarray}
where
\begin{eqnarray}
-G^0&=&
\Gamma\rho\kappa_0a_\mathrm{r}T_{g0}^4
-\Gamma\rho(\kappa_0-\sigma_0\Gamma^2\beta^2)E_\mathrm{r}
\nonumber\\
&&+\Gamma\rho[\kappa_0-\sigma_0(2\Gamma^2-1)]\beta_jF_\mathrm{r}^j
\nonumber\\
&&+\Gamma^3\rho\sigma_0\beta_j\beta_kP_\mathrm{r}^{jk},
\label{eq:G0}
\end{eqnarray}
and
\begin{eqnarray}
-G^i&=&
\Gamma\rho\kappa_0a_\mathrm{r}T_{g0}^4\beta^i
+\Gamma^3\rho\sigma_0E_\mathrm{r}\beta^i
\nonumber\\&&
-\Gamma\rho(\kappa_0+\sigma_0)F_\mathrm{r}^i
-2\Gamma^3\rho\sigma_0\beta^i\beta_jF^j_\mathrm{r}
\nonumber\\&&
+\Gamma\rho(\kappa_0+\sigma_0)\beta_jP_\mathrm{r}^{ij}
+\Gamma^3\rho\sigma_0\beta^i\beta_j\beta_kP^{jk}.
\label{eq:G1}
\end{eqnarray}

\subsection{Operator Splitting}
The numerical code integrates these equations by a similar way to those presented in \cite{2013ApJ...772..127T}. 
At first, we separate the basic equations into the advection and interaction parts. 
The former part deals with the evolution of hydrodynamical variables and the advection of radiation, both of which are governed by Equations (\ref{eq:continuity})-(\ref{eq:radiation_flux}) with $G^0=G^i=0$. 
The latter governs the exchange of energy and momenta between gas and radiation and is written as follows,
\begin{eqnarray}
\frac{\partial (\rho\Gamma)}{\partial t}&=&0
\label{eq:mass_conservation}\\
\frac{\partial( \rho h\Gamma^2\beta^i)}{\partial t}&=&G^i,
\label{eq:momentum2}
\\
\frac{\partial (\rho h\Gamma^2-P_\mathrm{g})}{\partial t}&=&G^0,
\label{eq:energy2}
\\
\frac{\partial E_\mathrm{r}}{\partial t}&=&-G^0,
\label{eq:Er}
\\
\frac{\partial F^i_\mathrm{r}}{\partial t}&=&-G^i.
\label{eq:Fr}
\end{eqnarray}
The numerical integration of the basic equations from $n$th step $t=t^n$ to the next step $t^{n+1}=t^n+\Delta t$ is realized by the so-called operator splitting technique, i.e., integrating the advection and interaction parts one after the other. 

The equations for hydrodynamics can be integrated by using some standard techniques for numerical hydrodynamics. 
We use the 2nd-order reconstruction scheme introduced by \cite{1977JCoPh..23..276V} to interpolate physical variables in each numerical cell and the HLLC scheme for special relativistic hydrodynamics \citep{2005MNRAS.364..126M} to calculate the numerical fluxes at the interface of two neighboring cells. 
In the following, we describe our method to numerically integrate the advection part for radiation and the interaction part. 

\subsection{Advection for Radiation}
The advection equations for the radiation energy density $E_\mathrm{r}$ and the flux $F^i_\mathrm{r}$ are integrated by using the so-called M1 closure scheme, which was first introduced by \cite{1984JQSRT..31..149L}. 
In order to integrate the advection equations for radiation, one needs to express the radiation pressure tensor in terms of $E_\mathrm{r}$ and $F^i_\mathrm{r}$. 
In the M1 closure scheme, the following formula is used,
\begin{equation}
P^{ij}_\mathrm{r}=\left(\frac{1-\chi}{2}\delta^{ij}+\frac{3\chi-1}{2}n^in^j\right)E_\mathrm{r},
\label{eq:M1}
\end{equation}
where $\delta^{ij}$ is the Kronecker's delta, the vector $n^i$ denotes the normalized radiation flux, 
\begin{equation}
n^i=\frac{F^{i}_\mathrm{r}}{|{\textbf{\textit {F}}}_\mathrm{r}|},
\end{equation}
and the variable $\chi$ is defined as a function of the flux normalized by the radiation energy density $f^i=F^i_\mathrm{r}/E_\mathrm{r}$,
\begin{equation}
\chi=\frac{3+4|{\textbf{\textit f}}|^2}{5+2\sqrt{4-3|{\textbf{\textit f}}|^2}}.
\end{equation}
From the definitions of the radiation energy density and flux, Equations (\ref{eq:def_er}) and (\ref{eq:def_fr}), the norm of the vector $f^i$ does not exceed unity, $|{\textbf{\textit f}}|\leq 1$. 
One can easily see that the parameter $\chi$ yields $1/3$ in the limit of an isotropic radiation field $f^i=0$, while the free-streaming limit, $|{\textbf{\textit f}}|=1$, leads to $\chi=1$, and these values correctly reproduce the expressions of the radiation pressure tensor, $P_\mathrm{r}^{ij}/E_\mathrm{r}=\delta^{ij}/3$ and $P^{ij}_\mathrm{r}/E_\mathrm{r}=n^in^j$, in these two limits. 
In addition, Equation (\ref{eq:M1}) is designed so that the expression is Lorentz invariant, which enables us to use the same expression in any inertial frame. 

Although we describe our method to integrate the advection part in one-dimensional cartesian coordinate for the sake of simplicity, the extension of the method to multi-dimension is straightforward. 
The radiation energy density and the radiative flux at $i$th cell at time $t=t^{n+1}$ can be calculated by the following way,
\begin{equation}
E_{\mathrm{r},i}(t^{n+1})=E_{\mathrm{r},i}(t^n)-\frac{\Delta t}{\Delta x}
\left({F}_{\mathrm{r},i+1/2}-{F}_{\mathrm{r},i-1/2}\right),
\end{equation}
and
\begin{equation}
F_{\mathrm{r},i}(t^{n+1})=F_{\mathrm{r},i}(t^n)-\frac{\Delta t}{\Delta x}
\left({P}_{\mathrm{r},i+1/2}-{P}_{\mathrm{r},i-1/2}\right),
\end{equation}
where ${F}_{\mathrm{r},i+1/2}$ and ${P}_{\mathrm{r},i+1/2}$ are the radiation flux and the radiation pressure tensor at the interface of the cell, which are evaluated by the well-known HLL scheme \citep{HLL}. 
In the scheme, the radiation energy density and the radiation flux averaged over the volume of a numerical cell are interpolated into the interface by using van Leer's scheme \citep{1977JCoPh..23..276V}. 
Thus, we obtain two sets of $E_\mathrm{r}$ and $F_\mathrm{r}$ at the interface $x=x_{i+1/2}$ interpolated from the cells at the left and right sides of the interface, which are denoted by $(E_{\mathrm{r},i+1/2,L},F_{\mathrm{r},i+1/2,L})$ and $(E_{\mathrm{r},i+1/2,R},F_{\mathrm{r},i+1/2,R}$), respectively. 
Furthermore, one obtains the values of the radiation pressure tensor, $P_{\mathrm{r},i+1/2,L}$ and $P_{\mathrm{r},i+1/2,R}$ from Equation (\ref{eq:M1}). 

The HLL scheme use the maximum and minimum speeds, $\lambda_+$ and $\lambda_-$, of waves expected to form at the interface where a couple of different radiation fields are in contact with each other. 
The numerical fluxes at the interface can be expressed in terms of the wave speeds and two sets of the variables, $(E_{\mathrm{r},i+1/2,L},F_{\mathrm{r},i+1/2,L},P_{\mathrm{r},i+1/2,L})$ and $(E_{\mathrm{r},i+1/2,R},F_{\mathrm{r},i+1/2,R},P_{\mathrm{r},i+1/2,R})$, as follows,
\begin{eqnarray}
{F}_{\mathrm{r},i+1/2}=&&
(\lambda_+-\lambda_-)^{-1}\left[
\lambda_+F_{\mathrm{r},i+1/2,L}+\lambda_-F_{\mathrm{r},i+1/2,R}
\right.
\nonumber\\&&
\left.+
\lambda_+\lambda_-(E_{\mathrm{r},i+1/2,R}-E_{\mathrm{r},i+1/2,L})
\right],
\end{eqnarray}
and
\begin{eqnarray}
{P}_{\mathrm{r},i+1/2}=&&(\lambda_+-\lambda_-)^{-1}
\left[\lambda_+P_{\mathrm{r},i+1/2,L}+\lambda_-P_{\mathrm{r},i+1/2,R}\right.
\nonumber\\&&+
\left.
\lambda_+\lambda_-(F_{\mathrm{r},i+1/2,R}-F_{\mathrm{r},i+1/2,L})\right].
\end{eqnarray}
One needs to calculate the maximum and minimum wave speeds appearing in the above equations. 
We follow the strategy adopted by \cite{2013MNRAS.429.3533S} and \cite{2013ApJ...772..127T}, in which the evaluation of the wave speeds for a cell changes according to the optical thickness, $\Delta \tau=\rho_0(\kappa_0+\sigma_0)\Delta x$, corresponding to the width of the cell. 
We basically assume that the maximum and the minimum wave speeds are given by the speed of light traveling into $\pm x$-direction, $\lambda_+=-\lambda_-=1$. 
However, for cells with a sufficiently large optical thickness, $\Delta \tau>100$,  we reduce the absolute values of the speeds by a factor of $4/(3\Delta \tau)$,
\begin{equation}
\lambda_+=\frac{4}{3\Delta \tau},\ \ \ 
\lambda_-=-\frac{4}{3\Delta \tau},
\end{equation}
which prevents the scheme from being too diffusive in highly optically thick media. 

The propagation of a photon ray in an optically thin media is solved as a test problem for the method outlined above. 
The setup and results of the calculation are described in detail in Appendix \ref{sec:beam_test}. 
We have confirmed that our treatment of the advection of radiation successfully reproduces the propagation of the photon rays. 

\subsection{Interaction between Radiation and Gas}
We transform Equations (\ref{eq:mass_conservation})-(\ref{eq:Fr}) into convenient forms. 
At first, we subtract the inner product of $\beta^i$ and Equation (\ref{eq:momentum2}) from Equation (\ref{eq:energy2}). 
After some algebraic manipulations, one obtains the following expression,
\begin{eqnarray}
&&\frac{E_\mathrm{g}+P_\mathrm{g}}{\Gamma}\frac{\partial \Gamma}{\partial t}
+\frac{\partial E_\mathrm{g}}{\partial t}
\\
&&\hspace{4em}=
-\rho\Gamma\kappa_0\left(
\frac{a_\mathrm{r}T_\mathrm{g}^4}{\Gamma^2}-E_\mathrm{r}+2\beta_jF^j_\mathrm{r}-\beta_j\beta_kP^{jk}_\mathrm{r}\right)
\equiv H^0.
\nonumber
\label{eq:interaction1}
\end{eqnarray}
Next, the subtraction of the $i$-component of Equation (\ref{eq:momentum2}) from the product of $\beta^i$ and Equation (\ref{eq:energy2}) yields
\begin{eqnarray}
&&\rho h\Gamma^2\frac{\partial\beta^i}{\partial t}-\beta^i\frac{\partial P_\mathrm{g}}{\partial t}
\\
&&\hspace{4em}=
-\rho\Gamma(\kappa_0+\sigma_0)(E_\mathrm{r}\beta^i-F^i_\mathrm{r}
-\beta^i\beta_jF^j_\mathrm{r}+\beta_jP^{ij}_\mathrm{r})
\nonumber\\
&&\hspace{4em}\equiv H^i,
\nonumber
\label{eq:interaction2}
\end{eqnarray}
Furthermore, the sum of Equations (\ref{eq:energy2}) and (\ref{eq:Er}) leads to the total energy conservation,
\begin{equation}
\frac{\partial (\rho h\Gamma^2-P_\mathrm{g}+E_\mathrm{r})}{\partial t}=0,
\end{equation}
while that of Equations (\ref{eq:momentum2}) and (\ref{eq:Fr}) yields the momentum conservation,
\begin{equation}
\frac{\partial( \rho h\Gamma^2\beta^i+F_\mathrm{r}^i)}{\partial t}=0.
\end{equation}
These four equations and the mass conservation (\ref{eq:mass_conservation}) are discretized in time to numerically integrate the interaction part of the basic equations. 

We integrate these equations implicitly. 
Therefore, the right-hand sides of Equations (\ref{eq:interaction1}) and (\ref{eq:interaction2}), $H^0$ and $H^i$, are evaluated by using variables at $t=t^{n+1}$. 
From the mass conservation, the product $U$ of the density and the Lorentz factor should be constant during the integration of the equations. 
Therefore, calculating the product $U$ from the density and the Lorentz factor at $t=t^n$, one obtains the following relation for the density $\rho^{n+1}$ and the Lorentz factor $\Gamma^{n+1}$ at $t=t^{n+1}$,
\begin{equation}
U=\rho^{n+1}\Gamma^{n+1}.
\label{eq:U}
\end{equation}
In similar ways, introducing the total momentum and energy, $S^i_\mathrm{tot}$ and $E_\mathrm{tot}$, which can be calculated from the variables at $t=t^n$ and should also be constant during the time step, the radiation energy density and the radiation flux at $t=t^{n+1}$, $E_\mathrm{r}^{n+1}$ and $F_\mathrm{r}^{i,n+1}$, are expressed in terms of $\beta^{i,n+1}$ and $P_\mathrm{g}^{n+1}$ as follow,
\begin{equation}
E_\mathrm{r}^{n+1}=E_\mathrm{tot}-U\Gamma^{n+1}-\left[
(\Gamma^{n+1})^2\frac{\gamma}{\gamma-1}-1\right]P_\mathrm{g}^{n+1}
\label{eq:En+1}
\end{equation}
\begin{eqnarray}
F_\mathrm{r}^{i,n+1}=&&
S^{i}_\mathrm{tot}-U\Gamma^{n+1}\beta^{i,n+1}
\\&&
-\frac{\gamma(\Gamma^{n+1})^2}{\gamma-1}\beta^{i,n+1}P_\mathrm{g}^{n+1}.
\nonumber
\label{eq:Fn+1}
\end{eqnarray}
On the other hand, the discretization of Equations (\ref{eq:interaction1}) and (\ref{eq:interaction2}) in time leads to
\begin{eqnarray}
&&\frac{E_\mathrm{g}^{n+1}+P_\mathrm{g}^{n+1}}{\Gamma^{n+1}}(\Gamma^{n+1}-\Gamma^n)
\\&&\hspace{4em}
+E_\mathrm{g}^{n+1}-E_\mathrm{g}^{n}=\Delta t H^{0,n+1},
\nonumber
\end{eqnarray}
and
\begin{eqnarray}
&&Uh^{n+1}\Gamma^{n+1}(\beta^{i,n+1}-\beta^{i,n})
\\&&\hspace{4em}
-\beta^{i,n+1}(P_\mathrm{g}^{n+1}-P_\mathrm{g}^n)=\Delta t H^{i,n+1}.
\nonumber
\end{eqnarray}
These equations are solved to obtain the velocity and the pressure of the gas at $t=t^{n+1}$, $\beta^{i,n+1}$ and $P_\mathrm{g}^{n+1}$, with which the density, the radiation energy density, and the radiation flux, $\rho^{n+1}$, $E_\mathrm{r}^{n+1}$, and $F_\mathrm{r}^{i,n+1}$, are obtained from Equations (\ref{eq:U}), (\ref{eq:En+1}), and (\ref{eq:Fn+1}). 

%%%%%%%%%%%%%%%%%%%%%%%
%%%    
%%%%%%%%%%%%%%%%%%%%%%%
\section{SIMULATIONS OF SUPERNOVA SHOCK BREAKOUT}
\subsection{Numerical Setups}
\subsubsection{Progenitor Star Model}
We carry out simulations of the supernova shock breakout in the explosion of a blue supergiant star. 
We adopt a progenitor star model provided by \cite{1988PhR...163...13N} and \cite{1988Natur.334..508S} in our simulations. 
This progenitor star model has been used for numerical studies trying to reproduce observations of SN 1987A \citep[e.g.,][]{1988A&A...196..141S,1990ApJ...360..242S} and thus it provides a fair comparison between our results and the earlier studies. 
The radius and the mass of the star at the pre-supernova stage are $3.5\times 10^{12}$ cm and $14.6\ M_\odot$.  
The star has its atmosphere composed of hydrogen, helium and heavy elements, whose mass fractions are $X=0.565$, $Y=0.430$, and $Z=0.005$, respectively. 
We assume a steady wind with a constant normalization of $\dot{M}/(4\pi v_\mathrm{w})=5\times 10^{11}\ \mathrm{g}\ \mathrm{cm}^{-1}$ outside the star, where $\dot{M}$ and $v_\mathrm{w}$ denote the mass-loss rate and the wind velocity. 
Thus, the escape velocity at the stellar surface $\sim 330$ km s$^{-1}$ leads to the mass-loss rate of $3.3\times 10^{-6}$ $M_\odot$ yr$^{-1}$. 
The mass of the wind in the computational domain is $5\times 10^{-8} M_\odot$, which is much smaller than the mass of the ejected matter. 
Once this circumstellar medium is illuminated by the shock breakout emission, the gas is immediately photoionized and heated to temperatures similar to the radiation temperature of the emission, $\sim 10^5-10^6$ K, making electron scattering the dominant opacity source. 
The optical depth of the wind for electron scattering yields $\sim 0.03$. 
Thus, the density of the ambient gas is so small that it hardly affects the propagation of the shock and radiation after the breakout.

The simulations are performed on two-dimensional spherical coordinates $(r,\theta)$ covering radial and angular ranges of $3\times 10^8$ cm $\leq r \leq$ $1.5\times 10^{13}$ cm and $0\leq\theta\leq\pi$ with $2048\times512$ numerical cells. 

\subsubsection{Energy Deposition}
An explosion energy $E_\mathrm{exp}$ is deposited during the first $\tau_\mathrm{exp}$ seconds of the simulations as thermal energy. 
Numerically, we increment the thermal energy density, $u_\mathrm{th}=P_\mathrm{g}/(\gamma-1)$ in the numerical cells adjacent to the inner boundary and with the angular coordinate $\theta$ at a rate of
\begin{equation}
\frac{du_\mathrm{th}}{dt}=
\frac{E_\mathrm{exp}}{\tau_\mathrm{exp}}\frac{1+a\cos(2\theta)}{(1-a/3)\Delta V},
\label{eq:u_th}
\end{equation}
where $\Delta V$ is the total volume of the numerical cells in which the explosion energy is deposited. 
Here we have introduced a parameter $a$ describing the deviation of the energy deposition from spherical symmetry. 
Integrating Equation (\ref{eq:u_th}) over the volume of the numerical cells adjacent to the inner boundary, one obtains the rate of the energy injection  into the whole numerical domain,
\begin{equation} 
\frac{E_\mathrm{exp}}{\tau_\mathrm{exp}}\int^{\pi}_0
\frac{1+a\cos(2\theta)}{2(1-a/3)}\sin\theta d\theta=\frac{E_\mathrm{exp}}{\tau_\mathrm{exp}},
\end{equation}
Thus, for a positive $a(\leq 1)$, $(1+a)/(1-a)$ times more energy is deposited at the symmetry axis than at the equatorial plane. 
We perform simulations with $a=0.0$, $0.2$, $0.5$, and $0.8$, corresponding to $(1+a)/(1-a)=1.0,\ 1.5$, $3.0$, and $9.0$, while the other free parameters are fixed, $E_\mathrm{exp}=10^{51}$ erg and $\tau_\mathrm{exp}=0.1$ s. 

\subsubsection{Opacities}
We assume that electron scattering and free-free absorption are the sources of scattering and absorption opacities. 
We use the following formulae,
\begin{equation}
\sigma_0=0.2(1+X)\ \mathrm{cm}^2\ \mathrm{g}^{-1}
\end{equation}
for electron scattering, and
\begin{equation}
\kappa_0=3.7\times10^{22}(1+X)(1-Z)\rho T^{-7/2}\ \mathrm{cm}^{2}\ \mathrm{g}^{-1}
\end{equation}
for free-free absorption \citep[see, e.g.,][]{1979rpa..book.....R}. 
We note that free-free emission is the only process to produce photons.

\subsection{Spherical Energy Deposition}
At first, we present results of the spherically symmetric model ($a=0.0$) and compare them with some of the earlier calculations with spherical symmetry. 
\subsubsection{Shock Propagation and Breakout}
Figure \ref{fig:snap_a0} shows the spatial distributions of the radiation energy density and the mass density at several epochs. 
At the very beginning of the simulation, a shocked region with high radiation energy density forms around the center of the progenitor star (top left panel). 
The shock wave propagates in the star (top right panel) and reaches the stellar surface at $t\sim 6.4\times 10^3$ s after the energy injection (middle left panel). 
Until the shock emergence from the stellar surface, the optical depth measured from the shock front to the stellar surface is sufficiently large, leading to the radiation front identical with the shock front as seen in the top panels of Figure \ref{fig:snap_a0}. 
Since the mean free path of photons in the ambient medium is much larger than the scale of the computational domain, radiation from the post-shock region can travel almost freely at the speed of light after the breakout (middle right and bottom left panels). 
As a result, the whole computational domain is filled with radiation with high energy density several hundreds seconds after the shock emergence (bottom right panel). 
The ejecta move at much slower velocities than the speed of light. 
Therefore, the radius of the ejecta at $t=7.0\times 10^3$ s is only slightly larger than the radius of the progenitor star. 
Although the simulation is performed in two-dimensional spherical coordinates, the ejecta and the radiation front keep spherical symmetry since the explosion energy is deposited in a spherical manner. 

%%%%%%%%%%%%%%%%%%%%%%%%%%%%%
\begin{figure*}[tbp]
\begin{center}
\includegraphics[scale=0.18]{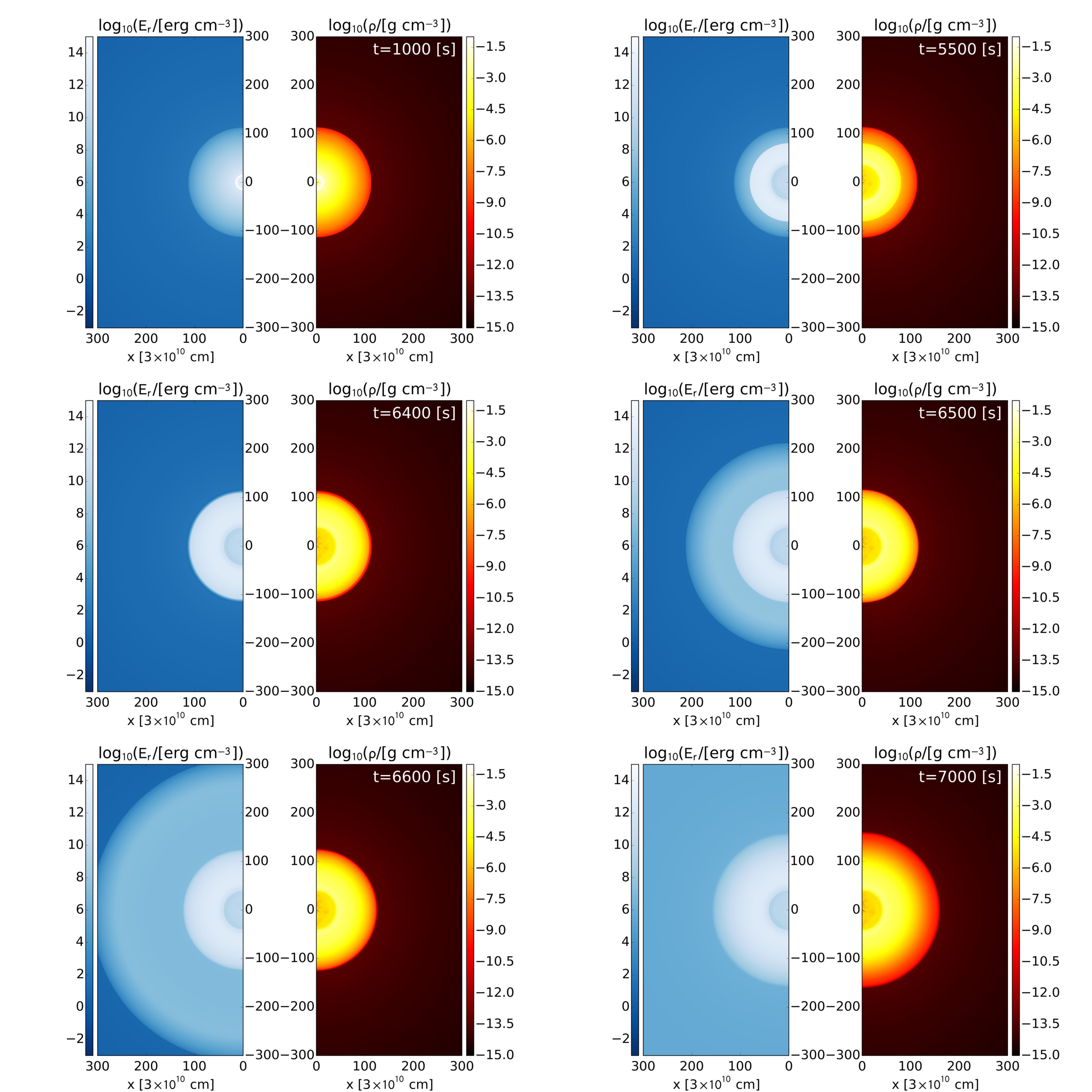}
\caption{Spatial distributions of the radiation energy density (left) and mass density (right) at several epochs for the spherical model $(a=0)$. 
The panels correspond to snapshots at $t=1.0\times 10^3$ (top left), $5.5\times 10^3$ (top right), $6.4\times 10^3$ (middle left), $6.5\times 10^3$ (middle right), $6.6\times 10^3$, (bottom left), and $7.0\times 10^3$ (bottom right) s. }
\label{fig:snap_a0}
\end{center}
\end{figure*}
%%%%%%%%%%%%%%%%%%%%%%%%%%%%%

The results are consistent with earlier calculations. 
For example, \cite{1990ApJ...360..242S} performed a radiation-hydrodynamic simulation using the same progenitor model as this work by employing the flux limited diffusion approach for radiative transfer. 
In their model with the explosion energy of $10^{51}$ erg, the shock wave reaches the stellar surface around 2 hours after the energy deposition. 
\cite{1992ApJ...393..742E} carried out similar calculations with variable Eddington factor method and showed that it took 1.9 hours after the energy deposition. 
Although the shock emergence in our simulation occurs slightly earlier, $t=1.8$ hours, than these studies, we conclude that our result is satisfactorily consistent with these earlier studies. 
%After the shock emergence, the interstellar space surrounding the ejecta is soon filled by radiation with an energy density of the order of $\sim 10^{8}$ erg cm$^{-3}$, which corresponds to an effective temperature of the order of $\sim 10^5$ K. 

\subsubsection{Breakout Emission}
%%%%%%%%%%%%%%%%%%%%%%%%%%%%%
\begin{figure*}[tbp]
\begin{center}
\includegraphics[scale=0.3]{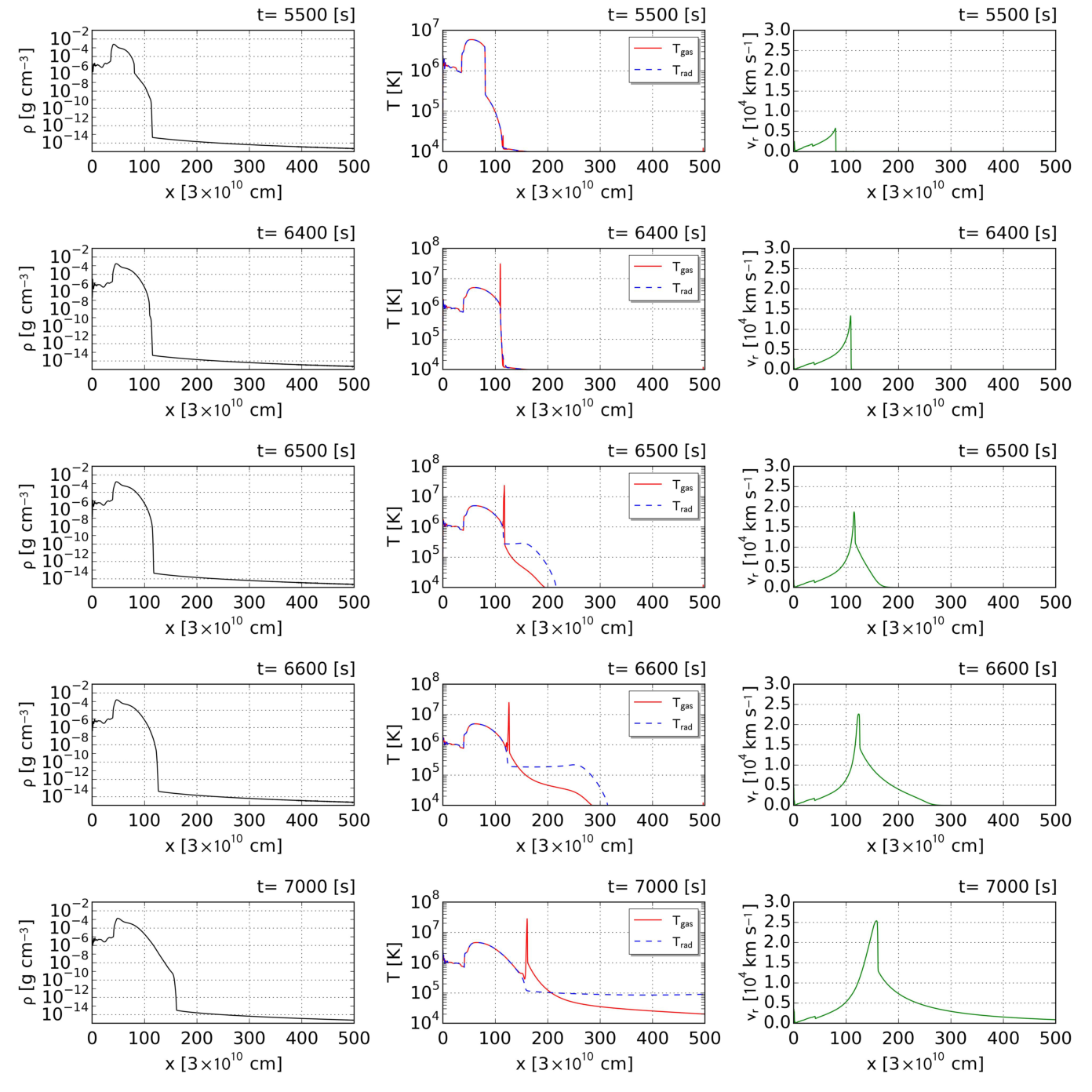}
\caption{Radial profiles of the density (left column), the gas and radiation temperatures (middle column), and the radial velocity (right column) at $t=5.5\times 10^3$, $6.4\times 10^3$, $6.5\times 10^3$, $6.6\times 10^3$, and $7.0\times 10^3$ s, from top to bottom. }
\label{fig:slice_a0}
\end{center}
\end{figure*}
%%%%%%%%%%%%%%%%%%%%%%%%%%%%%

The radial profiles presented in Figure \ref{fig:slice_a0} clearly demonstrate the evolution of the coupling between gas and radiation at the moment of the shock emergence. 
When the optical depth from the shock front to the stellar surface is still large ($t=5.5\times 10^3$ s), the radiation and gas are in equilibrium and the mixture behaves as a single fluid. 
However, as the shock front approaches the surface, the density at the shock front becomes smaller and thus the time required to maintain the equilibrium between radiation and gas becomes longer. 
As a result, the inefficient conversion of the internal energy of the shocked gas into radiation results in the deviation of the gas temperature from the radiation temperature ($t=6.4\times 10^3$ s). 
The gas temperature starts deviating from the radiation temperature when the pre-shock density becomes as small as $\rho_\mathrm{bo}\simeq 10^{-9}$ $\mathrm{g\ cm}^{-3}$ as shown in the density profile at $t=6.4\times 10^3$ s in Figure \ref{fig:slice_a0}. 
When the shock front reaches this layer, the post-shock velocity is $v_\mathrm{bo}\simeq 1.3\times 10^9$ cm s$^{-1}$. 
Thus, the post-shock temperature can be estimated as follows,
\begin{equation}
T_\mathrm{bo}=\left[\frac{(\gamma+1)\rho_\mathrm{bo}v_\mathrm{bo}^2}{2(\gamma-1)a_\mathrm{r}}\right]^{1/4}\simeq 10^6\ \mathrm{K},
\label{eq:Tbo}
\end{equation}
under the assumption that the post-shock gas still maintains the gas-radiation equilibrium and the post-shock internal energy is dominated by radiation. 
The radiation temperature obtained in the numerical simulation at the moment of the shock emergence ($t=6.4\times 10^3$ s) well agrees with this value. 
After the shock emergence, the radiation front propagates in the interstellar space and heats the ambient gas. 
The velocity of the outermost layer reaches a terminal value of $v_\mathrm{max}=2.5\times 10^4$ km s$^{-1}$ about 500 s after the breakout, which is consistent with \cite{1992ApJ...393..742E}. 

Finally, we mention the color temperature of the breakout emission. 
The value derived above, Equation (\ref{eq:Tbo}), is the temperature of the layer above which the shocked gas cannot create photons so efficiently that the gas-radiation equilibrium is realized. 
Therefore, most of thermal photons escaping as the shock breakout emission should be created in regions below this layer and the color temperature of the emission reflects this temperature $T_\mathrm{bo}\sim 10^6$ K, while the effective temperature of the emission is $(2$-$3)\times 10^{5}$ K as shown in the radial profiles of the radiation temperature in Figure \ref{fig:slice_a0}. 
This discrepancy between the color and effective temperatures has also been reported by \cite{1992ApJ...393..742E}.

\subsection{Bipolar Energy deposition}

%%%%%%%%%%%%%%%%%%%%%%%%%%%%%
\begin{figure*}[tbp]
\begin{center}
\includegraphics[scale=0.18]{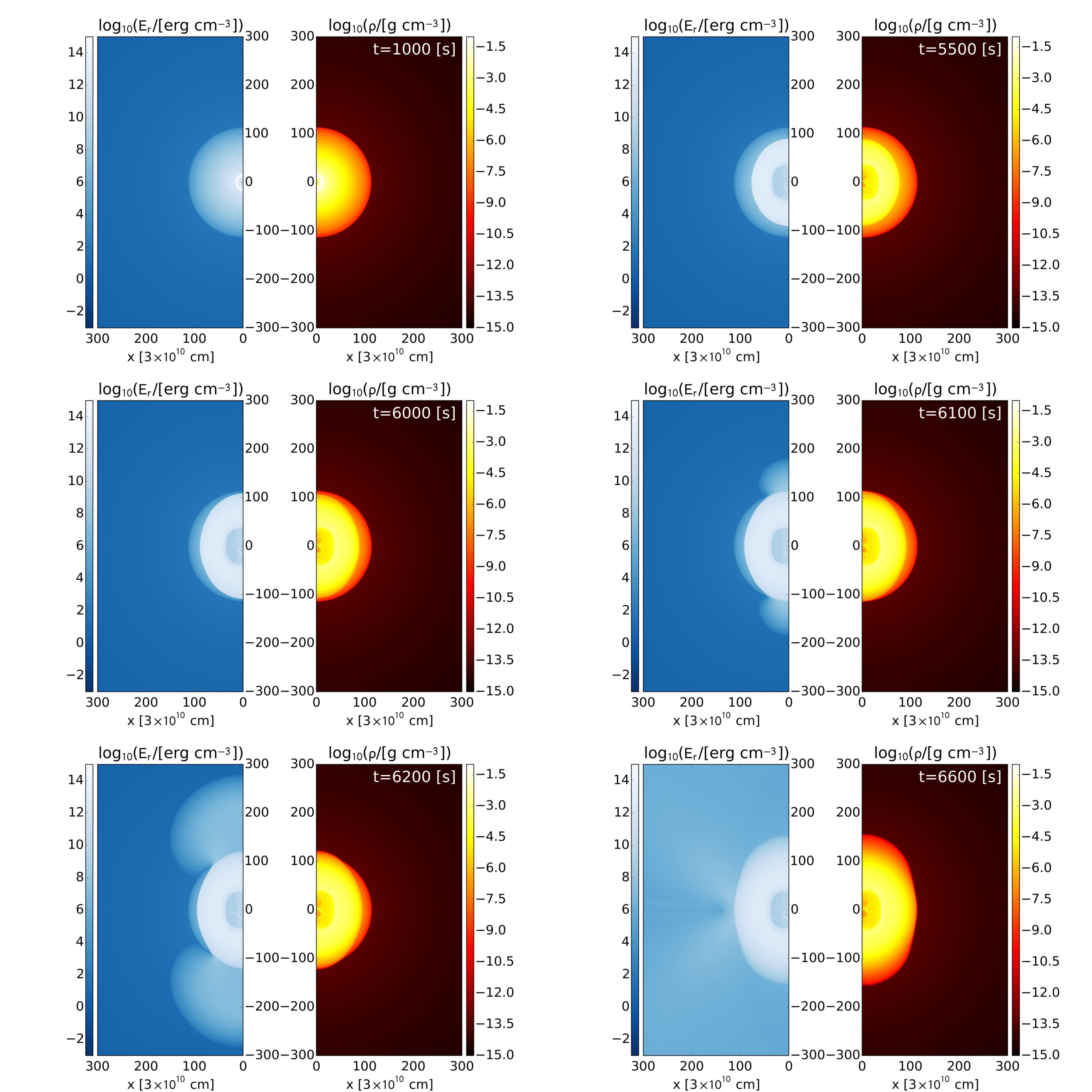}
\caption{Same as Figure \ref{fig:snap_a0}, but for the model with $a=0.5$. 
The panels correspond to snapshots at $t=1.0\times 10^3$ (top left), $5.5\times 10^3$ (top right), $6.0\times 10^3$ (middle left), $6.1\times 10^3$ (middle right), $6.2\times 10^3$, (bottom left), and $6.6\times 10^3$ (bottom right) s. }
\label{fig:snap_a5}
\end{center}
\end{figure*}
%%%%%%%%%%%%%%%%%%%%%%%%%%%%%

When the explosion energy is deposited in an aspherical manner, the shock breakout occurs in a different way. 
Figure \ref{fig:snap_a5} presents results of the model with $a=0.5$. 
In this case, the shock wave propagates faster along the symmetry axis and slower near the equatorial plane, resulting in the shocked region elongated in the direction of the symmetry axis as seen in the top right panel of Figure \ref{fig:snap_a5}. 
As a result, the shock emerges from the poles of the star at first (middle left panel). 
Then, the radiation trapped in the shocked region starts escaping into the ambient space from the poles (middle right and bottom left panels). 
After the shock emergence is completed, the ejecta are expected to gradually approach free expansion. 

Figure \ref{fig:slice_a5} shows the radial profiles of the density, the gas and radiation temperatures, and the radial velocity along angular coordinates of $\theta=0^\circ$, $30^\circ$, $45^\circ$, $60^\circ$, and $90^\circ$ at the moment when the shock wave propagating in each radial direction emerges from the surface. 
In this model, the shock emergence takes $\Delta t\simeq 500$ seconds from the beginning at $t=6.05\times 10^3$ s until the shock propagating near the equatorial plane reaches the surface at $t=6.55\times 10^3$ s, which makes the aspherical shock breakout different from the spherical one. 
The radial profiles at $\theta=0^\circ$ (top panel) are similar to those in the spherical model when the decoupling between radiation and gas starts at $t=6.4\times 10^3$ s. 
On the other hand, in radial profiles at large $\theta$, a region with high radiation and gas temperatures appears in front of the shock front prior to the shock emergence along the radial direction. 
This is a consequence of the multi-dimensional effects also shown in Figure \ref{fig:snap_a5}. 
In other words, the radiation emitted around the poles of the star fills the region surrounding the star and heat the circumstellar medium before the breakout occurs around the equatorial plane. 

Despite the time delay of the shock emergence, the density and the velocity at the shock front along different radial directions take similar values when radiation and gas start decoupling. 
The temperature profiles at different angles also look similar to each other except for the preheating region ahead of the shock front. 
Thus, we expect emission with similar color temperatures from different parts of the stellar surface. 
The basic properties of the shock breakout realized in the aspherical models, such as the radiation energy, the ejecta velocity, the color temperature, are similar to those in the spherical model. 
We consider the effect of the aspherical property of the shock propagation on the light curve of the breakout emission in the next section.

%%%%%%%%%%%%%%%%%%%%%%%%%%%%%
\begin{figure*}[tbp]
\begin{center}
\includegraphics[scale=0.3]{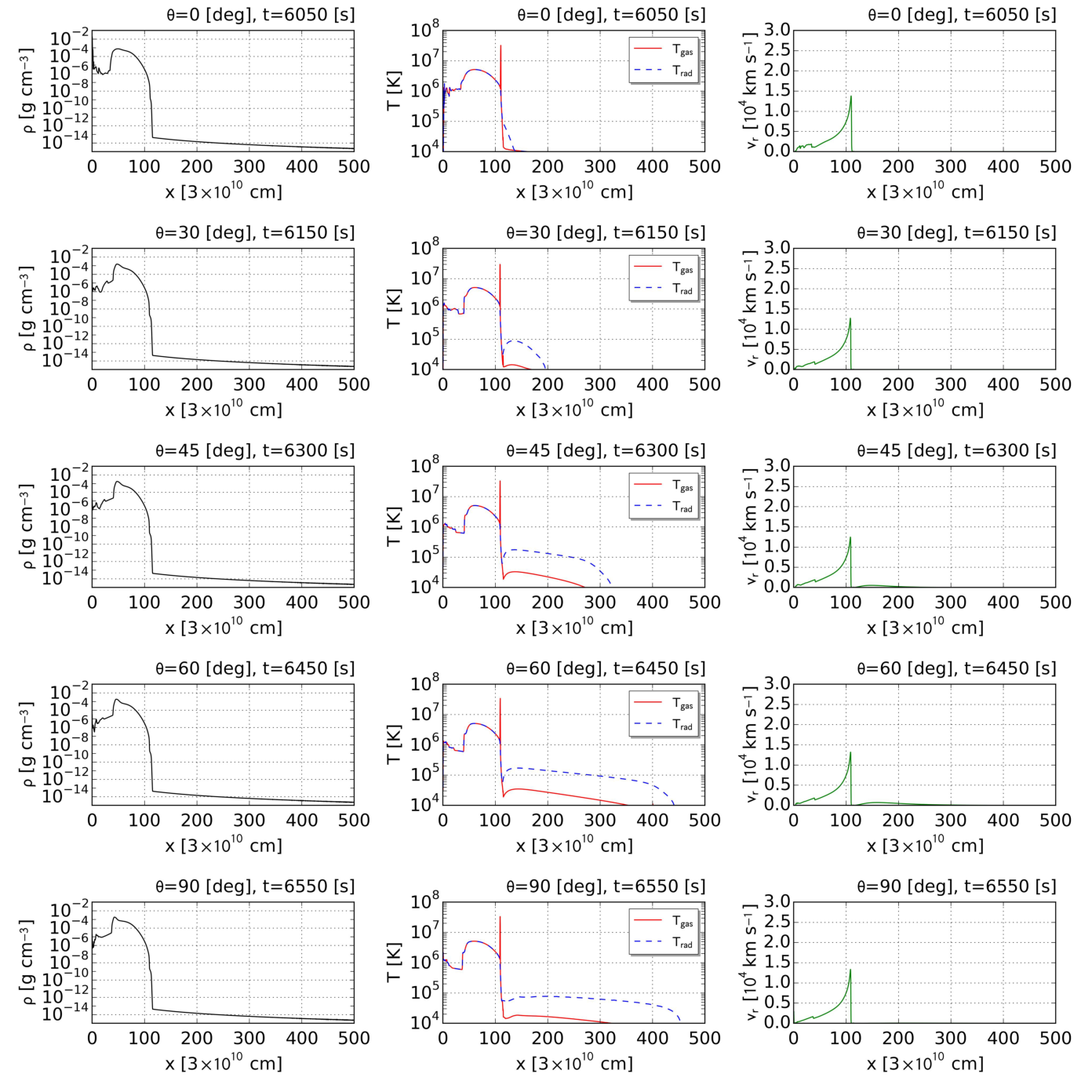}
\caption{Radial profiles of the density (left column), the gas and radiation temperatures (middle column), and the radial velocity (right column) along various radial directions. 
The panels show the profiles at which the shock wave propagating in each radial direction emerges from the surface for $\theta=0^\circ$, $30^\circ$, $45^\circ$, $60^\circ$, and $90^\circ$, from top to bottom.}
\label{fig:slice_a5}
\end{center}
\end{figure*}
%%%%%%%%%%%%%%%%%%%%%%%%%%%%%

%%%%%%%%%%%%%%%%%%%%%%%
%%%    
%%%%%%%%%%%%%%%%%%%%%%%
\section{GEOMETRICAL EFFECTS ON SHOCK BREAKOUT LIGHT CURVES}
\subsection{Derivation of light curves}
The light curves of the shock breakout emission are calculated by using the following ray-tracing method. 

When the intensity $I_\nu$ in the laboratory frame is given as a function of the spatial coordinates ${\textbf{\textit{x}}}$, the time $t$, the frequency $\nu$, and the direction ${\textbf{\textit{l}}}$, the transfer equation is described as follows,
\begin{eqnarray}
\frac{\partial I_{\nu}}{\partial t}+({\textbf{\textit{l}}}\cdot\nabla)I_{\nu}=&&
{\cal D}^{-1}\alpha_{\nu'}'[{\cal D}^3B'_{\nu'}(T_\mathrm{g})-I_\nu]
\nonumber\\&&
+{\cal D}^{-1}\sigma_{\nu'}'({\cal D}^3J_{\nu'}'-I_\nu),
\end{eqnarray}
where $B'_{\nu'}(T_\mathrm{g})$ and $J'_{\nu'}$ denote a Planck function with a gas temperature $T_\mathrm{g}$ and the mean intensity per unit frequency in the comoving frame of the flow. 
The factor ${\cal D}$ denotes the so-called Doppler factor, which describes the difference in the frequencies in the laboratory and comoving frames, $\nu$ and $\nu'$,
\begin{equation}
\nu=\Gamma(1-\beta_il^i)^{-1}\nu'\equiv {\cal D}\nu'. 
\end{equation}
Photons with a frequency $\nu'$ are absorbed and scattered at rates given by the coefficients $\alpha_{\nu'}'$ and $\sigma_{\nu'}'$, which are also given in the comoving frame. 

Integrating the transfer equation with respect to the frequency, one obtains the following equation governing the temporal evolution of the frequency-integrated intensity $I$,
\begin{eqnarray}
\frac{\partial I}{\partial t}+({\textbf{\textit{l}}}\cdot\nabla)I=&&
{\cal D}^{-1}\alpha'\left({\cal D}^4\frac{\sigma_\mathrm{SB}T_\mathrm{g}^4}{\pi}-I\right)
\nonumber\\
&&+{\cal D}^{-1}\sigma'\left({\cal D}^4\frac{E_\mathrm{r}'}{4\pi}-I\right),
\end{eqnarray}
where $\alpha'$ and $\sigma'$ denote the frequency-averaged values of the coefficients and $\sigma_\mathrm{SB}$ is the Stefan-Boltzmann constant. 
The radiation energy density $E_\mathrm{r}'$ is defined in the comoving frame. 
Thus, the Lorentz transformation of the quantities, $E_\mathrm{r}$, $F_\mathrm{r}^i$, and $P^{ij}_\mathrm{r}$, which are defined in the laboratory frame and used in our simulations, gives this value, 
\begin{equation}
E_\mathrm{r}'=\Gamma^2\left(E_\mathrm{r}-2\beta_iF_\mathrm{r}^i+\beta_i\beta_jP_\mathrm{r}^{ij}\right).
\end{equation}

Here we define the source function $S'$ in the comoving frame in the following way, 
\begin{equation}
S'=(\alpha'+\sigma')^{-1}\left(
\alpha'\frac{\sigma_\mathrm{SB}T_\mathrm{g}^4}{\pi}+\sigma'\frac{E_\mathrm{r}'}{4\pi}\right).
\end{equation}

Introducing the source function and frequency-averaged absorption and scattering coefficients in the laboratory frame as follows, $S={\cal D}^4S'$, $\alpha={\cal D}^{-1}\alpha'$, and $\sigma={\cal D}^{-1}\sigma'$, 
the frequency-integrated transfer equation is rewritten in the following form,
\begin{equation}
\frac{\partial I}{\partial t}+({\textbf{\textit{l}}}\cdot\nabla)I=
(\alpha+\sigma)(S-I),
\label{eq:transfer_eq}
\end{equation}
which is solved to obtain the temporal evolution of the bolometric luminosity. 

We consider a distant observer seeing the explosion from a viewing angle $\Theta_\mathrm{v}$ with respect to the symmetry axis. 
The configuration considered here is schematically represented in the left panel of Figure \ref{fig:projection}. 
The observer sees photons escaping from the ejecta into the direction specified by a vector ${\textbf{\textit{l}}}_\mathrm{v}=(\sin\Theta_\mathrm{v},0,\cos\Theta_\mathrm{v})$ in the $x$-$z$ plane. 
To calculate the luminosity of the emission, we consider an imaginary plane (referred to as ``screen'') perpendicular to the direction vector ${\textbf{\textit{l}}}_\mathrm{v}$, which is located at a distance $D$, from the center and define two-dimensional spherical coordinates ($r_\mathrm{v},\phi$) on the screen, whose origin is at ${\textbf{\textit{x}}}=D{\textbf{\textit{l}}}_\mathrm{v}$.

At first, we integrate the transfer equation (\ref{eq:transfer_eq}) along a photon ray parallel to the vector ${\textbf{\textit{l}}}_\mathrm{v}$ and passing through a point ($r_\mathrm{v},\phi$) on the screen by using snapshots of a simulation. 
The position of the point on the screen can be expressed in terms of the coordinates $(r_\mathrm{v},\phi)$ as follows 
\begin{equation}
{\textbf{\textit x}}_\mathrm{sc}=D{\textbf{\textit l}}_\mathrm{v}+{\textbf{\textit n}}(r_\mathrm{v},\phi),
\end{equation}
where the vector in the second term of the right-hand side is given by,
\begin{equation}
{\textbf{\textit n}}(r_\mathrm{v},\phi)=
(\cos\Theta_\mathrm{v}\cos\phi,
\sin\phi,
-\sin\Theta_\mathrm{v}\cos\phi).
\end{equation} 
Thus, the location of a point at a distance $s$ from ${\textbf{\textit x}}={\textbf{\textit x}}_\mathrm{sc}$ along the ray is specified by ${\textbf{\textit x}}={\textbf{\textit x}}_\mathrm{sc}-{\textbf{\textit l}}_\mathrm{v}s$. 
From the transfer equation, it is straightforward to obtain the intensity at ${\textbf{\textit x}}_\mathrm{sc}-s{\textbf{\textit l}}_\mathrm{v}$ from that at ${\textbf{\textit x}}_\mathrm{sc}-(s+\Delta s){\textbf{\textit l}}_\mathrm{v}$ with small $\Delta s$,
\begin{eqnarray}
I(t-s,{\textbf{\textit x}}_\mathrm{sc}-s{\textbf{\textit l}}_\mathrm{v})=&&
I(t-s-\Delta s,{\textbf{\textit x}}_\mathrm{sc}-(s+\Delta s){\textbf{\textit l}}_\mathrm{v})e^{-\tau}
\nonumber\\&&
+S(1-e^{-\tau}),
\end{eqnarray}
where the optical depth along the ray is given by,
\begin{equation}
\tau=(\alpha+\sigma)\Delta s.
\end{equation}
Here we have assumed that the source function $S$ and the coefficients $\alpha$ and $\sigma$ are constant in a small duration and distance $\Delta s$, and that their values are evaluated at $t-s-\Delta s$. 
This formula is repeatedly used along the ray to obtain the intensity on the screen at time $t$ from a series of snapshots obtained from a radiation-hydrodynamics simulation. 
We note that the same opacities as the radiation-hydrodynamics simulations are used in the ray-tracing calculations.

%%%%%%%%%%%%%%%%%%%%%%%%%%%%%
\begin{figure*}[tbp]
\begin{center}
\includegraphics[scale=0.4]{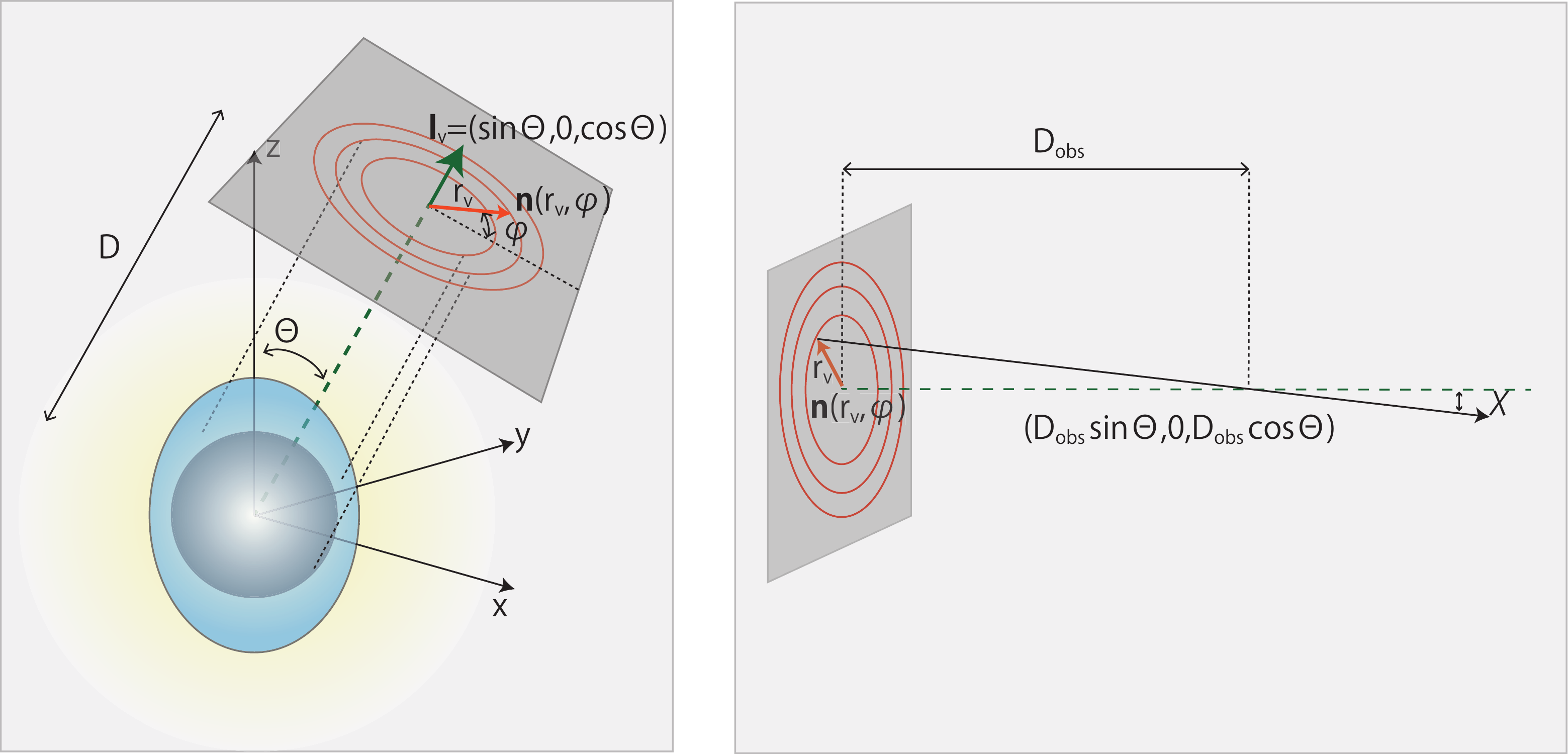}
\caption{Schematic view of the configuration considered in the ray-tracing treatment in calculating the light curve of the shock breakout emission. 
Photons traveling into the direction specified by the direction vector ${\textbf{\textit{l}}}_\mathrm{v}$ are considered. 
The photon rays intersect with a plane perpendicular to the direction vector, which is shown as a gray region in the figure. }
\label{fig:projection}
\end{center}
\end{figure*}
%%%%%%%%%%%%%%%%%%%%%%%%%%%%%

Next, we derive a formula to calculate the light curve of the emission, which can be used for both bolometric and frequency dependent cases. 
We consider photon rays traveling from various points on the screen into the location of an observer ${\textbf{\textit{x}}}_\mathrm{obs}=(D_\mathrm{obs}\sin\Theta,0,D_\mathrm{obs}\cos\Theta)$ at a distance $D_\mathrm{obs}$ much larger than the scale of the screen, as shown in the right panel of Figure \ref{fig:projection}. 
The intensity of a photon ray passing through a point $(r_\mathrm{v},\phi)$ on the screen is denoted by $I'(r_\mathrm{v},\phi)$. 
The flux $F_\mathrm{obs}$ of the radiation at ${\textbf{\textit{x}}}_\mathrm{obs}$ is obtained by integrating the intensity of the photon rays multiplied by the direction cosine $\cos\chi$ measured with respect to the direction vector ${\textbf{\textit{l}}}_\mathrm{v}$ over the solid angle,
\begin{equation}
F_\mathrm{obs}=\int I'\cos\chi d\Omega.
\end{equation} 
Here, the infinitesimal solid angle element in the above equation can be expressed in terms of the coordinates $(r_\mathrm{v},\phi)$ as follows,
\begin{equation}
d\Omega=\frac{r_\mathrm{v}dr_\mathrm{v}d\phi}{D_\mathrm{obs}^2}.
\end{equation}
Taking the limit $D_\mathrm{obs}\rightarrow \infty$, the angle $\chi$ approaches $0$ and thus the intensity $I'(r_\mathrm{v},\phi)$ is now identical with $I(t,{\textbf{\textit{x}}}_\mathrm{sc})$. 
Furthermore, since the isotropic luminosity $L$ at the point ${\textbf{\textit{x}}}_\mathrm{obs}$ is given by $L=4\pi D_\mathrm{obs}^2F_\mathrm{obs}$, the luminosity at the limit $D_\mathrm{obs}\rightarrow \infty$ can be written as follows,
\begin{equation}
L(t)=4\pi\int I(t,{\textbf{\textit{x}}}_\mathrm{sc})r_\mathrm{v}dr_\mathrm{v}d\phi.
\end{equation}
This equation gives the luminosity of the shock breakout emission as a function of time $t$. 

\subsection{Light traveling time effect}

Performing the ray-tracing calculation, we obtain the bolometric light curve of the shock breakout emission. 
In Figure \ref{fig:LC_a0}, the light curve calculated by the ray-tracing method for the spherical model is shown. 
One can also calculate the radiation energy lost per unit time through the outer boundary of the numerical domain $r=R_\mathrm{out}$ as follows,
\begin{equation}
L_\mathrm{emit}(t)=2\pi R_\mathrm{out}^2\int F_r(t,R_\mathrm{out},\theta) d\cos\theta,
\label{eq:Lemit}
\end{equation}
where $F_r$ is the radiation flux at time $t$ along a radial direction at an angular coordinate $\theta$. 
Since this value corresponds to the energy leaving the numerical domain through a sphere with the outer radius $R_\mathrm{out}$ covering the progenitor star, the evolution of $L_\mathrm{emit}(t)$ is different from the light curve of the emission seen by a distant observer. 
The temporal evolution of the emitted power $L_\mathrm{emit}$ is compared with the light curve  obtained from the ray-tracing method in Figure \ref{fig:LC_a0}, which clearly demonstrates that the ray-tracing correction is necessary. 

Since our progenitor model has the pre-supernova radius of $R_\ast=3.5\times 10^{12}$ cm, the light curve after the ray-tracing reflects its light crossing time, $R_\ast/c\simeq 10^2$ s. 
This is the reason why the early bright part of the bolometric light curve lasts for about $10^2$ s. 
In other words, the light curve of the shock breakout emission is smeared out with the time scale given by the light traveling time of the stellar radius, which is called the light traveling effect. 
The bolometric luminosity in Figure \ref{fig:LC_a0} reaches $L=2.5\times10^{44}$ erg s$^{-1}$ at the peak. 
\cite{1992ApJ...393..742E} reported the peak luminosity without the light traveling time correction of $L=6\times10^{44}$ erg s$^{-1}$ and showed that the light travel time effect reduces the value by a factor of $\sim2$ to $L=(3.6-3.8)\times 10^{44}$ erg s$^{-1}$. 
Thus, the peak luminosity in our model matches their value within a factor of 2. 
%Furthermore, our bolometric light curve after the ray-tracing correction shows good agreement with that in \cite{1992ApJ...393..742E}. 
\cite{1990ApJ...360..242S} reported the peak luminosity of $L=4\times 10^{44} $ erg s$^{-1}$ without the light traveling time correction. 
Thus, assuming the reduction of the peak luminosity by a factor of $\sim 2$, our peak luminosity well agrees with theirs. 
Therefore, we again confirm that our spherical model satisfactorily reproduces the earlier studies with spherical symmetry. 

%%%%%%%%%%%%%%%%%%%%%%%%%%%%%
\begin{figure}[tbp]
\begin{center}
\includegraphics[scale=0.18]{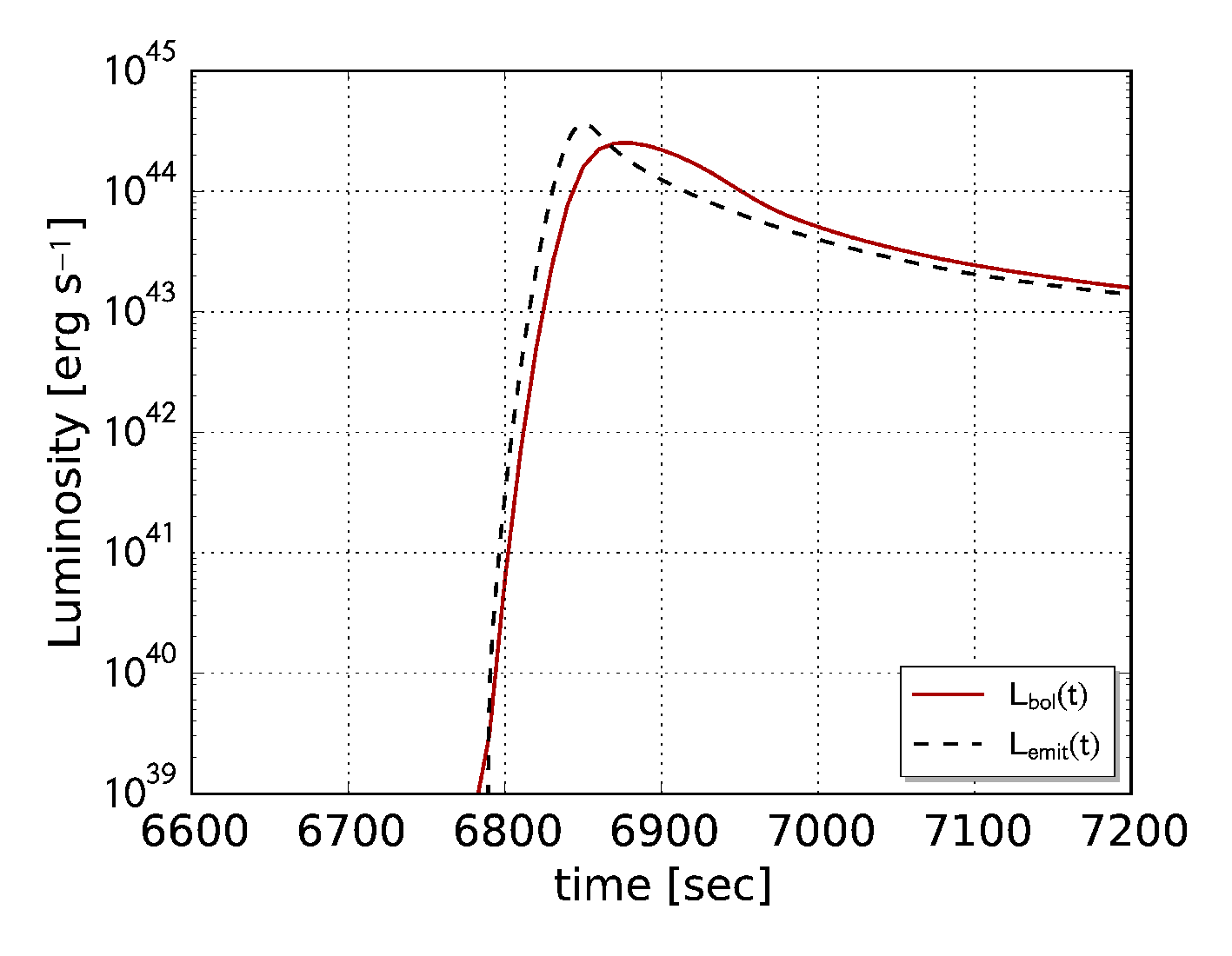}
\caption{Bolometric light curve of the model with $a=0$ (solid line). The rate of the loss of the radiation energy from the outer boundary calculated by Equation (\ref{eq:Lemit}) is also plotted as a dashed line. }
\label{fig:LC_a0}
\end{center}
\end{figure}
%%%%%%%%%%%%%%%%%%%%%%%%%%%%%

\subsection{Light curves of aspherical shock breakout}

%%%%%%%%%%%%%%%%%%%%%%%%%%%%%
\begin{figure}[tbp]
\begin{center}
\includegraphics[scale=0.25]{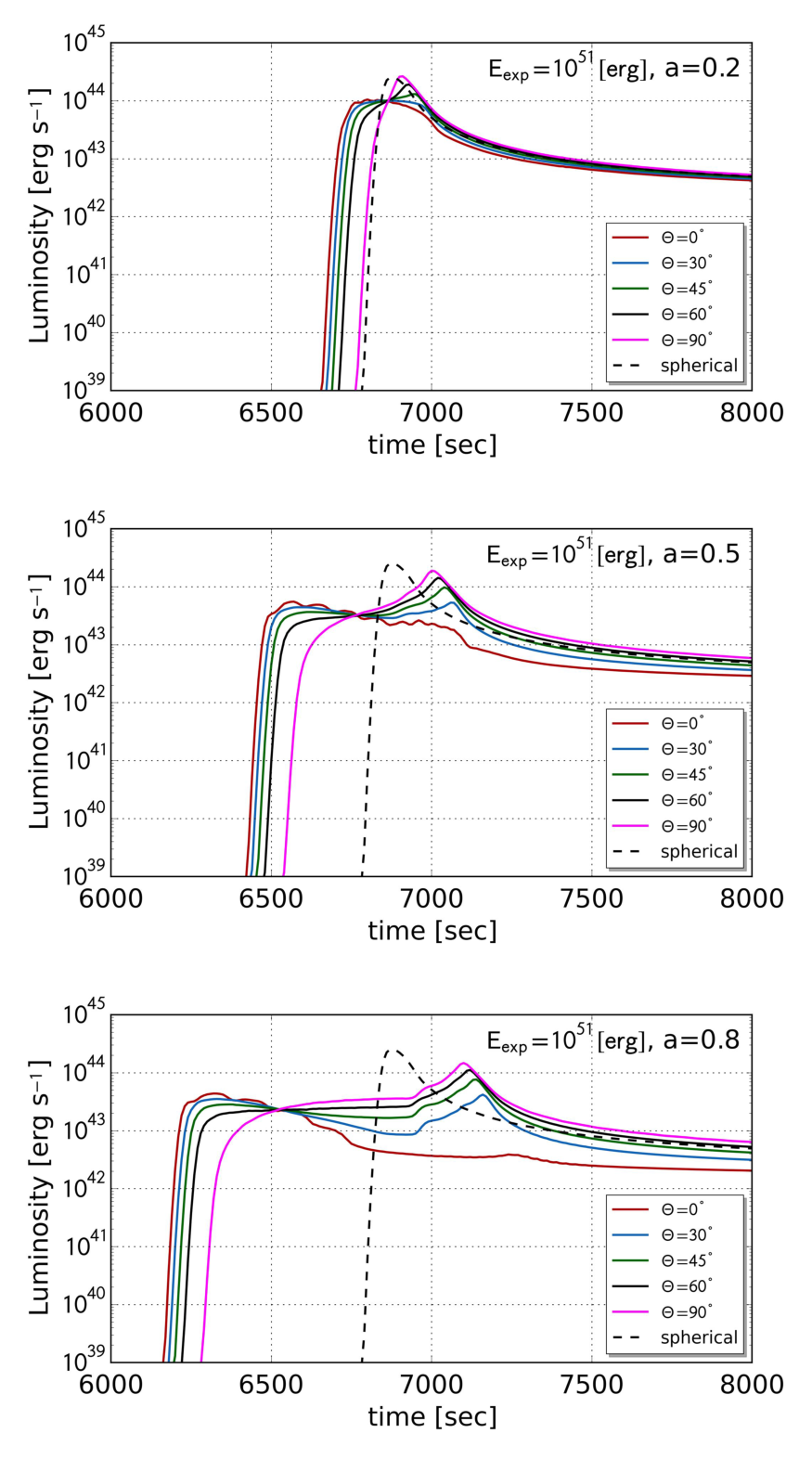}
\caption{Bolometric light curves of the models with $a=0.2$ (top panel), $0.5$ (middle panel), and $0.8$ (bottom panel) from the viewing angles of $\Theta=0^\circ$, $30^\circ$, $45^\circ$, $60^\circ$, and $90^\circ$. 
In each panel, the bolometric light curve of the spherical model ($a=0$) shown in Figure \ref{fig:LC_a0} is also plotted as a dashed line. }
\label{fig:light_curves}
\end{center}
\end{figure}
%%%%%%%%%%%%%%%%%%%%%%%%%%%%%

\subsubsection{Governing time scales}
In Figure \ref{fig:light_curves}, we show the bolometric light curves of the models with $a=0.2$, $0.5$, and $0.8$ observed from the viewing angles of $\Theta=0^\circ$, $30^\circ$, $45^\circ$, $60^\circ$, and $90^\circ$. 
In each panel, they are compared with that of the spherical model. 
We should note that the light curve with $\Theta=0^\circ$ suffers from artificial fluctuations, which is produced when the line of sight is in parallel with the symmetry axis. 

At first, we focus on the model with $a=0.5$ (the middle panel of Figure \ref{fig:light_curves}). 
The peak luminosities are smaller and the early bright emission lasts longer than the spherical model. 
These properties are explained by the geometrical effects. 
In general, models with aspherical energy deposition take longer time than the spherical model to complete the shock emergence due to the elliptic shape of the shock front. 
In the model with $a=0.5$, for example, the shock waves propagating along $\theta=0^\circ$ and $90^\circ$ reach the surface at $t=6050$ and $6550$ s  as shown in Figure \ref{fig:slice_a5}, leading to a duration of the shock emergence of $\Delta t\sim 500$ s. 
This makes the bolometric light curve of the model different from the spherical model. 
In the spherical model, the shock wave propagating along each radial direction simultaneously deposits a fraction of the shock kinetic energy into the stellar envelope as a thermal energy, which is rapidly converted to the radiation energy. 
As we have mentioned in the previous subsection, the light curve reflects the light traveling time of the stellar radius in this case. 
On the other hand, the energy deposition by the shock passage gradually occurs in aspherical cases, 
%In the model with $a=0.5$, it takes $\Delta t\simeq 500$ s for the shock emergence to be completed, which is longer than the light traveling time, $R_\ast/c$. 
while the amount of the radiation energy deposited by the shock passage in the aspherical models is similar to the spherical case. 
Thus, in the model with $a=0.5$, the radiation energy similar to the spherical model is released for a longer duration, making the earlier part of the bolometric light curve less luminous and longer-lived than that in the spherical model. 
This prolongation and modification of the earlier part of the bolometric light curve have been partly pointed out by our previous work \citep{2010ApJ...717L.154S}. 

However, there are some differences between the present and previous works. 
At first, the previous work was based on hydrodynamic simulations without radiative transfer and thus it used a greatly simplified model for the shock breakout emission. 
After the emergence of a shock traveling along each radial direction, the corresponding part of the stellar surface was assumed to emit blackbody emission for a fixed time, which was determined from the diffusion time of the breakout emission, and then it was shut off. 
This situation is qualitatively different from our radiation-hydrodynamic simulations, in which the shocked gas having emerged from the surface continues to produce emission even after the diffusion time. 
Thus, the sudden shut-off of the breakout emission adopted in the previous work is not appropriate. 
Furthermore, due to the limited resolution and the way of the energy injection different from this work, the delay between the shock emergence at the symmetry axis and on the equatorial plane in the previous work is smaller than those realized in this work, resulting in a significant difference in the time scale of the breakout emission. 

The later part of the bolometric light curves of different models in Figure \ref{fig:light_curves} (from $t=7.5\times 10^3$ s to $t=8.0\times 10^3$ s) look similar to each other. 
The later part of the shock breakout emission reflects the photospheric emission from the ejecta produced as a result of the shock emergence and cooling via adiabatic expansion. 
Thus, the rate of the decline of the bolometric luminosity should be governed by the expansion rate of the ejecta, which is given by the traveling time of the stellar radius by the maximum velocity of the ejecta, $R_\ast/v_\mathrm{max}$. 
Since the maximum velocities realized in the spherical and aspherical models are similar, $v_\mathrm{max}=2.5\times 10^4$ km s$^{-1}$, the corresponding decline rates should also be similar. 

\subsubsection{Dependence on the viewing angle}

Light curves viewed from large viewing angles are characterized by a gradually increasing  luminosity and a prominent peak before the luminosity starts declining ($t\sim 7.0\times 10^3$ s for the model with $a=0.5$). 
This peak is created due to emission from shocked gas around the equatorial plane, where shock waves propagating in different radial directions reach the surface nearly at the same time. 
The amount of radiation energy released from the stellar surface within a time interval is roughly proportional to the area of the surface hit by the shock wave during the interval, because the radiative flux on the stellar surface at the moment of the breakout does not strongly depend on the angular coordinate. 
This leads to a larger amount of radiation energy released per unit time from the stellar surface around the equatorial plane than around the symmetry axis. 

Another key ingredient to determine the brightness of the emission is the projected area of the emitting surface with respect to the line of sight. 
While a larger amount of radiation energy is released on the surface area close to the equatorial plane, the projected area of the emitting region near the equatorial plane is roughly proportional to $\cos(\pi/2-\Theta)$, which can make the emission less luminous for small viewing angles. 
As a result of the reduction, the bolometric light curve with $\Theta=0^\circ$ brightens at first and gradually declines without any prominent peak before entering the adiabatic cooling phase.

\subsubsection{Dependence on the parameter $a$}
In Figure \ref{fig:light_curves}, the bolometric light curves for the models with $0.2$ and $0.8$ are also shown. 
For models with higher degrees of asphericity, the deviation from the spherical case becomes more significant. 
Especially, for the model with $a=0.8$ (bottom panel of Figure \ref{fig:light_curves}), the initial bright phase lasts for up to $\sim 10^3$ s, which is about one order of magnitude longer than that in the spherical case. 
On the other hand, the model with $a=0.2$ exhibits small deviation from the spherical case. 
This behavior clearly demonstrates that highly aspherical explosions can produce shock breakout emission with longer duration. 
%Therefore, the degree of asphericity in the energy deposition in the star could be probed by the bolometric light curves of their shock breakout emission. 

%%%%%%%%%%%%%%%%%%%%%%%
%%%    DISCUSSIONS
%%%%%%%%%%%%%%%%%%%%%%%
\section{CONCLUSIONS AND DISCUSSIONS}\label{discussion}
In this study, we develop a two-dimensional radiation-hydrodynamics code and carry out simulations of supernova shock breakout from a blue supergiant progenitor with spherical and aspherical energy deposition. 
Our spherical model successfully reproduces the coupling of radiation and gas, the dynamical evolution of the shock and the ejecta, and the bolometric light curve of the shock breakout investigated by earlier studies with spherical symmetry. 
Furthermore, our aspherical models clarify effects of aspherical energy deposition on the dynamics of the shock emergence from the stellar surface and the bolometric light curve. 
In this section, we summarize the properties of the bolometric light curve from aspherical shock breakout and the prospects for constraining the explosion geometry from future observations of supernova shock breakout. 

\subsection{Implications for Observations of Supernova Shock Breakout}
When the explosion energy is deposited in a spherical way, the shock waves propagating in various radial directions simultaneously reach the stellar surface. 
Thus, the radiation energy deposited into the stellar envelope is expected to escape into the surrounding space within the diffusion time. 
In this case, the early bright part of the shock breakout emission lasts for the light traveling time of the stellar radius. 
On the other hand, in aspherical explosions, the time from the initiation to the end of the shock emergence can be longer than the light traveling time of the stellar radius, leading to the gradual release of the radiation energy having been deposited into the stellar envelope. 
In our simulations with $a=0.2$, $0.5$, and $0.8$, the total amount of radiation energy in the stellar envelope are similar to the spherical model. 
Therefore, the aspherical models show slowly-evolving and under-luminous bolometric light curves compared with the spherical model. 

The later part of the bolometric light curves shows similar decline rates for both spherical and aspherical models. 
After the shock emergence, the stellar material swept by the blast wave creates an expanding hot gas, which rapidly cools by adiabatic expansion. 
The later part corresponds to emission from this adiabatically cooling fireball, in which the decline rate of the bolometric luminosity is governed by the traveling time of the stellar radius by the maximum velocity of the ejecta, $R_\ast/v_\mathrm{max}$. 
As long as the same amount of the explosion energy is deposited and the deviation from the spherical symmetry is moderate as assumed in this study, the resulting supernova ejecta follow a similar dynamical evolution. 

Generally, photometric and spectroscopic observations of a CCSN at around days to months after the shock breakout provide us information on the typical velocity $v_\mathrm{ej}$ of the ejecta. 
Thus, the detection of the cooling tail of the shock breakout emission provides a clue to determine the stellar radius $R_\ast$, as suggested by earlier works and have already been applied to particular events \citep{2013Natur.494...65O,2014Natur.509..471G,2014ApJ...789..104O}. 
Furthermore, our findings imply that the early bright part of the bolometric light curve can be used as a tracer of aspherical shock fronts. 
If the duration of the early part of an observed bolometric light curve is found to be longer than the light crossing time of the stellar radius estimated from the cooling tail of the breakout emission, the present results indicate that the aspherical energy deposition may have prolonged the initial phase of the shock breakout. 

Recent observations have found some long-lived bright emission associated with the birth of CCSNe \citep{2008Natur.453..469S,2008ApJ...683L.131G,2010ApJ...724.1396O,2015ApJ...804...28G}. 
When it is interpreted as the shock breakout emission, it requires large stellar radii. 
It is suggested that an optically thick wind or an extended envelope should be present when the core collapse onsets, prolonging the early bright emission. 
In our simulations, the model with $a=0.8$ produces the early bright emission nearly an order of magnitude longer than the light crossing time of the progenitor radius, suggesting that the geometrical effects might be responsible for the observed long-lived emission. 
However, the adopted parameter $a=0.8$ indicates that $(1+a)/(1-a)=9$ times more energy is deposited at the symmetry axis than at the equatorial plane. 
In other words, highly aspherical energy deposition is required to account for the observed long-lived early emission by the multi-dimensional effect. 
However, it is uncertain whether such highly aspherical energy deposition could be commonly realized. 
From an observational point of view, it is important to clarify how common CCSNe with long-lived early emission are. 
The manner in which the explosion energy is deposited at gas around the collapsing iron core is also unclear. 
While the explosion energy is injected from the inner boundary as thermal energy in our simulation, it may be in the form of kinetic energy. 
In that case, the relation between the parameter $a$ and the resultant light curves wound be different from our results shown in this work. 

Detecting early emission from CCSNe from optical to UV or X-ray energy ranges have been paid a great attention and an interesting target of future missions, such as, ULTRASAT \citep{2014AJ....147...79S}. 
The geometry of the energy deposition and the blast wave in collapsing massive stars could be investigated through their early light curves of the shock breakout in the near future.

\subsection{Some Remarks}
Finally, we note some remarks on our results. 
At first, we have assumed that the gas is fully ionized and free-free absorption and electron scattering are the only processes absorbing and scattering photons. 
These assumptions are appropriate at the shock emergence, at which the gas temperature is of the order of $10^{6}$ K and thus the resultant bolometric light curves obtained by our simulations successfully reproduce that of \cite{1992ApJ...393..742E}, who used more sophisticated opacities. 
However, when the gas temperature of the ejecta becomes small, $T<10^{5}$ K, due to free expansion, free electrons start recombining to ions, which reduces electron scattering opacity, and bound-free process starts contributing to the total opacity. 
As shown by \cite{1992ApJ...393..742E}, the enhanced absorption and emission of photons lead to the formation of a cooling shell, which is not seen in our simulations. 
This difference clearly suggests that more sophisticated opacities should be used to investigate the dynamical evolution of the ejecta at later epochs. 

Next, our code treats frequency-integrated equations of radiation-hydrodynamics. 
Recent studies investigating the structure of radiation-mediated shocks suggest that post-shock gas temperature becomes much higher than that expected in the equilibrium between radiation and gas, which produces high energy photons in X-ray and gamma-ray energy ranges. 
This process cannot be treated by our code due to the grey approximation. 
However, even when such high energy photons are produced in a thin layer behind the shock front, the geometrical effects revealed in this study should play a critical role in determining the X-ray or gamma-ray light curve. 
 
\acknowledgments
We are grateful to an anonymous referee for his/her constructive comments on the manuscript.
A.K. thanks H. R. Takahashi for his helpful comments on the development of the radiation-hydrodynamics code. 
A.K. also thanks C. Matzner for constructive discussion on our results. 
Numerical calculations were in part carried out on the XC30 system at Center for Computational Astrophysics, National Astronomical Observatory of Japan. 
K.M. acknowledges financial support by JSPS KAKENHI Grant Number 26800100. The work by K.M. is partly supported
by WPI Initiative, MEXT, Japan.

\appendix
\section{Test problems for radiation hydrodynamics code}
In this section, results of several test problems carried out by our code are presented. 

\subsection{Beam test}\label{sec:beam_test}
At first, we carry out simulations of radiative transfer in two-dimensional cartesian coordinates. 
In this test problem, called ``beam test'', the propagation of a ray of photons in an optically thin medium is considered. 
The numerical domain is a two-dimensional plane in cartesian coordinates $(x,y)$, both of which range from $-100$ to $100$ in cm, covered by $512\times512$ numerical cells. 
The domain is initially filled with isotropic radiation field at a radiation temperature $T_\mathrm{i}=10^6$ K,
\begin{equation}
E_\mathrm{r}(x,y)=a_\mathrm{r}T_\mathrm{i}^4,\ \ \ 
F_\mathrm{r}^x(x,y)=F_\mathrm{r}^y(x,y)=0. 
\end{equation}
We do not consider any process absorbing and emitting radiation by the gas in the domain. 
We inject a photon ray from the left boundary by imposing the following boundary conditions at $x=-100$ cm,
\begin{eqnarray}
E_\mathrm{r}(-100,y)&=&a_\mathrm{r}T_\mathrm{b}^4,\nonumber\\
F_\mathrm{r}^x(-100,y)&=&a_\mathrm{r}T_\mathrm{b}^4\cos(\pi/4),\nonumber\\
F_\mathrm{r}^y(-100,y)&=&a_\mathrm{r}T_\mathrm{b}^4\sin(\pi/4),\ \ \ \mathrm{for}\ -80\ \mathrm{cm}\leq y\leq -60\ \mathrm{cm}. 
\end{eqnarray}
In Figure \ref{fig:beam_test}, some snapshots of the spatial distribution of the radiation energy density are shown. 
Form the left boundary, the photon ray is injected and propagate at the speed of light into the direction at the angle of $45^\circ$ with respect to the $x$-axis. 

\subsection{Shadow test}\label{sec:shadow_test}
The second test problem is called ``shadow test'' \citep{2003ApJS..147..197H}, in which an optically thick medium is surrounded by optically thin gas at rest in the same two-dimensional physical space as the beam test. 
We set the following initial condition for the density of the gas,
\begin{equation}
\rho_0(x,y)=\left\{\begin{array}{ccl}
1.0\ \mathrm{g\ cm}^{-3}&\mathrm{for}&\sqrt{x^2+y^2}<20\ \mathrm{cm}\\
10^{-5}\ \mathrm{g\ cm}^{-3}&&\mathrm{otherwise}
\end{array}\right.,
\end{equation}
while the velocity and the pressure of the gas are set to zero because we only treat the evolution of the radiation field and does not solve those of hydrodynamical variables. 
The medium absorbs photons at a constant opacity $\kappa_\mathrm{a}=1.0\ \mathrm{cm}^2\ \mathrm{g}^{-1}$ and no scattering process is considered. 
The radiation is injected from the left boundary, at which the following conditions are imposed,
\begin{eqnarray}
E_\mathrm{r}(-100,y)&=&a_\mathrm{r}T_\mathrm{b}^4,\nonumber\\
F_\mathrm{r}^x(-100,y)&=&a_\mathrm{r}T_\mathrm{b}^4,\nonumber\\
F_\mathrm{r}^y(-100,y)&=&0,\ \ \ \ \ \ \ \ \mathrm{for} -100\ \mathrm{cm}\leq y\leq 100\ \mathrm{cm},
\end{eqnarray}
with $T_\mathrm{b}=10^8$ K. 

The spatial distributions of the radiation energy density at several epochs are shown in Figure \ref{fig:shadow_test}. 
Immediately after the injection of the radiation, the radiation front propagates toward the right direction at the speed of right. 
When the front reaches to the optically thick medium at $x=-20$ cm, the medium starts absorbing photons whose rays intersect the medium. 
The "shadow" of the medium appears as a result of the absorption of the radiation as seen in the lower right panel of Figure \ref{fig:shadow_test}, proving that the propagation and the absorption of radiation are correctly solved.

\subsection{Relativistic shock tube problem}
Finally, we solve shock tube problems treating the coupling between gas and radiation in a radiative shock \citep{2008PhRvD..78b4023F}, which are commonly used as a test for relativistic radiation-hydrodynamics codes \citep{2012MNRAS.426.1613R,2013MNRAS.429.3533S,2013ApJ...764..122T,2013ApJ...772..127T}. 
A plane parallel numerical domain from $x=-40$ to $x=40$ is filled with a couple of media with different conditions, which are initially separated at $x=0$. 
We perform four simulations with different sets of the initial condition (test 1, 2, 3, and 4) as in \cite{2008PhRvD..78b4023F}. 
Thus, the values of the physical variables corresponding to the right and left states, the adiabatic index, the absorption coefficient, and the normalized radiation constant are exactly same as those given in \cite{2008PhRvD..78b4023F}. 
The temporal evolutions of the flows in test 1, 2, 3, and 4 are followed until $t=5000$, $100$, $20$, and $100$, and the profiles of the velocity, density, pressure, radiation energy density, and the flux, are shown in Figure \ref{fig:shock_tube}. 
The radiation energy density and the flux in the comoving frame are plotted, while the velocity, the density, and the pressure are defined in the laboratory frame. 

Although \cite{2008PhRvD..78b4023F} derived analytical solutions of the problems under the Eddington approximation, in which the Eddington tensor in the comoving frame is given by $P^{ij}/E_\mathrm{r}=\delta ^{ij}/3$, the corresponding solutions adopting the M1 closure scheme are not known. 
Furthermore, as demonstrated by \cite{2013ApJ...772..127T}, results obtained by employing the Eddington approximation and the M1 closure scheme show slightly different profiles of physical quantities especially around the shock front. 
Thus, we compare our results with those of other authors' works adopting the M1 closure scheme rather than directly compare them with the analytical solutions provided by \cite{2008PhRvD..78b4023F}. 
We have confirmed that our code successfully reproduces results reported by \cite{2013ApJ...772..127T}.

%%%%%%%%%%%%%%%%%%%%%%%%%%%%%
\begin{figure}[tbp]
\begin{center}
\includegraphics[scale=0.3]{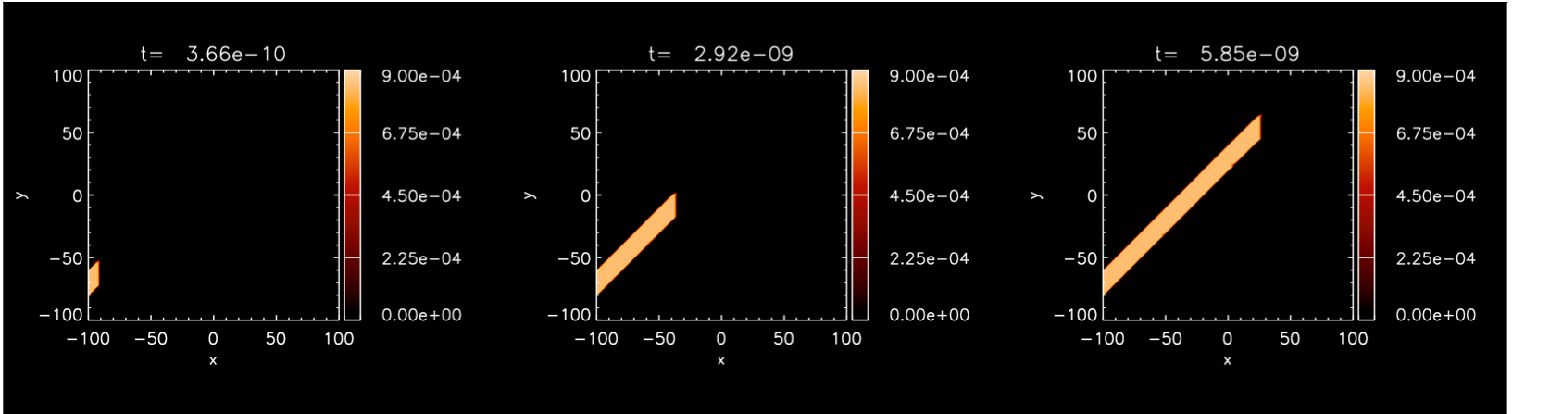}
\caption{Results of the beam test. 
The color-coded spatial distributions of the radiation energy density at several epochs are shown. }
\label{fig:beam_test}
\end{center}
\end{figure}
%%%%%%%%%%%%%%%%%%%%%%%%%%%%%

%%%%%%%%%%%%%%%%%%%%%%%%%%%%%
\begin{figure}[tbp]
\begin{center}
\includegraphics[scale=0.3]{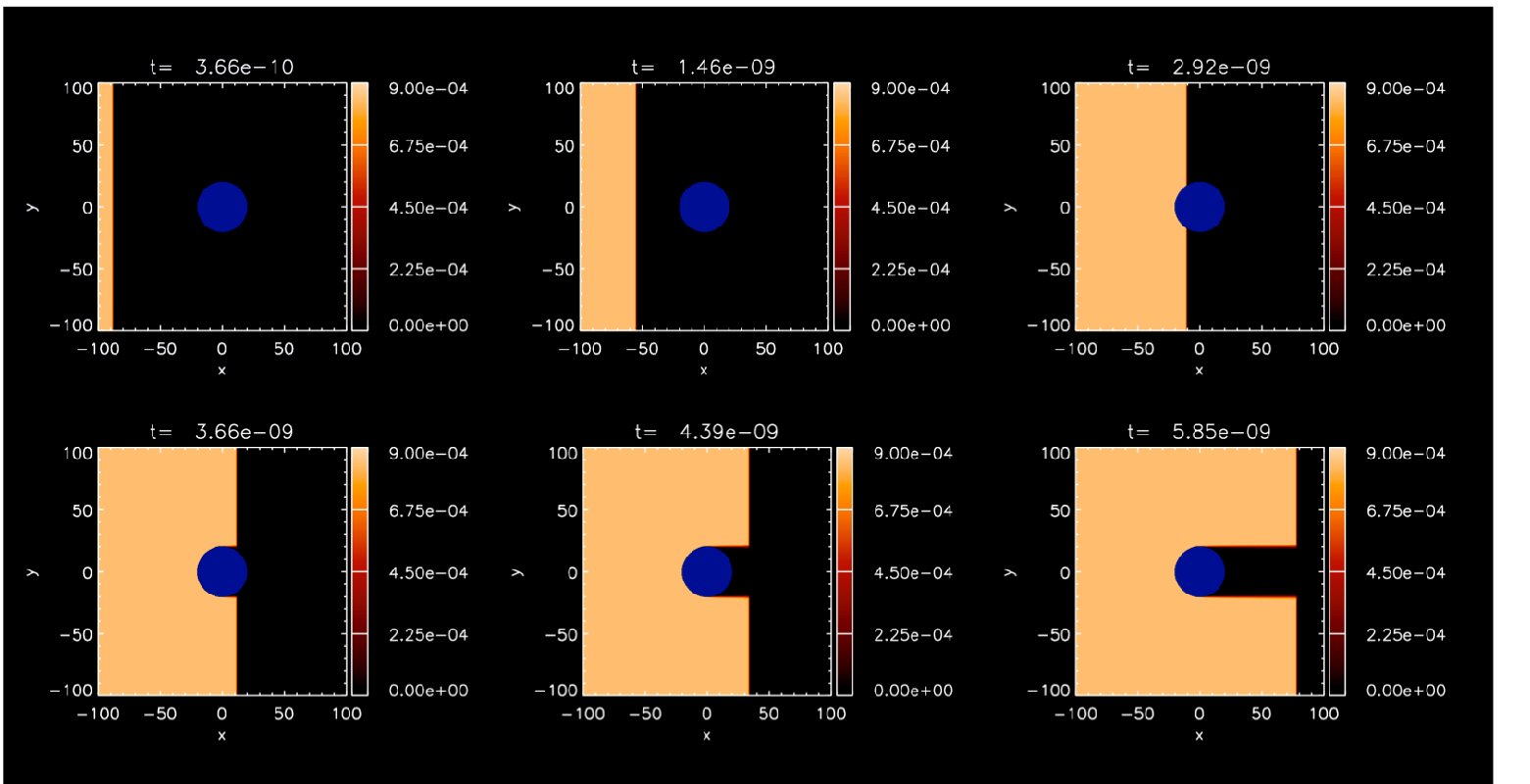}
\caption{Results of the shadow test. 
The color-coded spatial distributions of the radiation energy density at several epochs are shown.}
\label{fig:shadow_test}
\end{center}
\end{figure}
%%%%%%%%%%%%%%%%%%%%%%%%%%%%%

%%%%%%%%%%%%%%%%%%%%%%%%%%%%%
\begin{figure}[tbp]
\begin{center}
\includegraphics[scale=0.3]{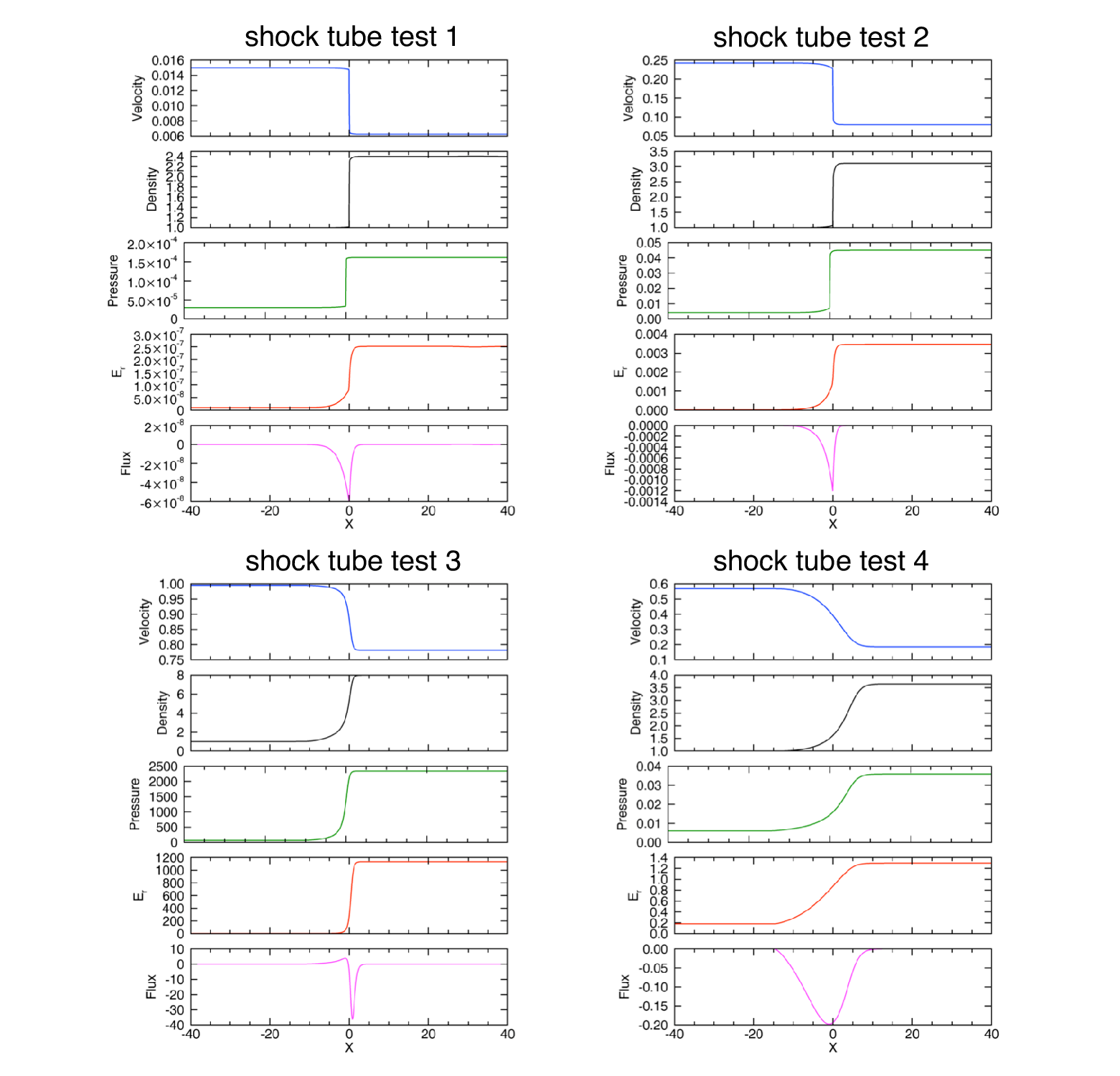}
\caption{Results of the shock tube tests. In each panel, the profiles of the velocity, density, pressure, the radiation energy density, and the radiation flux at the end of the simulation are plotted. }
\label{fig:shock_tube}
\end{center}
\end{figure}
%%%%%%%%%%%%%%%%%%%%%%%%%%%%%


\begin{thebibliography}{}
\bibitem[Balberg 
\& Loeb(2011)]{2011MNRAS.414.1715B} Balberg, S., \& Loeb, A.\ 2011, \mnras, 414, 1715 
\bibitem[Blinnikov et al.(2000)]{2000ApJ...532.1132B} Blinnikov, S., 
Lundqvist, P., Bartunov, O., Nomoto, K., 
\& Iwamoto, K.\ 2000, \apj, 532, 1132 
\bibitem[Budnik et al.(2010)]{2010ApJ...725...63B} Budnik, R., Katz, B., 
Sagiv, A., \& Waxman, E.\ 2010, \apj, 725, 63 
\bibitem[Campana et al.(2006)]{2006Natur.442.1008C} Campana, S., Mangano, 
V., Blustin, A.~J., et al.\ 2006, \nat, 442, 1008 
\bibitem[Chevalier(1992)]{1992ApJ...394..599C} Chevalier, R.~A.\ 1992, 
\apj, 394, 599
\bibitem[Chevalier 
\& Fransson(2008)]{2008ApJ...683L.135C} Chevalier, R.~A., \& Fransson, C.\ 2008, \apjl, 683, L135 
\bibitem[Chevalier 
\& Irwin(2011)]{2011ApJ...729L...6C} Chevalier, R.~A., \& Irwin, C.~M.\ 2011, \apjl, 729, L6 
\bibitem[Colgate(1974)]{1974ApJ...187..333C} Colgate, S.~A.\ 1974, \apj, 
187, 333
\bibitem[Couch et al.(2011)]{2011ApJ...727..104C} Couch, S.~M., Pooley, D., 
Wheeler, J.~C., \& Milosavljevi{\'c}, M.\ 2011, \apj, 727, 104
\bibitem[Ensman 
\& Burrows(1992)]{1992ApJ...393..742E} Ensman, L., \& Burrows, A.\ 1992, \apj, 393, 742 

%\bibitem[Ensman(1994)]{1994ApJ...424..275E} Ensman, L.\ 1994, \apj, 424, 275 
\bibitem[Falk 
\& Arnett(1973)]{1973ApJ...180L..65F} Falk, S.~W., \& Arnett, W.~D.\ 1973, \apjl, 180, L65
\bibitem[Falk 
\& Arnett(1977)]{1977ApJS...33..515F} Falk, S.~W., \& Arnett, W.~D.\ 1977, \apjs, 33, 515
\bibitem[Falk(1978)]{1978ApJ...225L.133F} Falk, S.~W.\ 1978, \apjl, 225, 
L133
\bibitem[Farris et al.(2008)]{2008PhRvD..78b4023F} Farris, B.~D., Li, 
T.~K., Liu, Y.~T., \& Shapiro, S.~L.\ 2008, \prd, 78, 024023
\bibitem[Gal-Yam et al.(2014)]{2014Natur.509..471G} Gal-Yam, A., Arcavi, 
I., Ofek, E.~O., et al.\ 2014, \nat, 509, 471 
\bibitem[Gezari et al.(2008)]{2008ApJ...683L.131G} Gezari, S., Dessart, L., 
Basa, S., et al.\ 2008, \apjl, 683, L131 
\bibitem[Gezari et al.(2015)]{2015ApJ...804...28G} Gezari, S., Jones, 
D.~O., Sanders, N.~E., et al.\ 2015, \apj, 804, 28 
\bibitem[Ginzburg 
\& Balberg(2014)]{2014ApJ...780...18G} Ginzburg, S., \& Balberg, S.\ 2014, \apj, 780, 18 
\bibitem[Ginzburg 
\& Balberg(2012)]{2012ApJ...757..178G} Ginzburg, S., \& Balberg, S.\ 2012, \apj, 757, 178
\bibitem[Gonz{\'a}lez et 
al.(2007)]{2007A&A...464..429G} Gonz{\'a}lez, M., Audit, E., \& Huynh, P.\ 2007, \aap, 464, 429
\bibitem[Harten et al.(1983)]{HLL}
Harten, A., Lax, P., and van Leer, B.\ 1983, SIAM review, 25, 35
\bibitem[Hayes 
\& Norman(2003)]{2003ApJS..147..197H} Hayes, J.~C., \& Norman, M.~L.\ 2003, \apjs, 147, 197
\bibitem[Hayes et al.(2006)]{2006ApJS..165..188H} Hayes, J.~C., Norman, 
M.~L., Fiedler, R.~A., et al.\ 2006, \apjs, 165, 188 
\bibitem[Janka et al.(2007)]{2007PhR...442...38J} Janka, H.-T., Langanke, 
K., Marek, A., Mart{\'{\i}}nez-Pinedo, G., 
M\"{u}ller, B.\ 2007, \physrep, 442, 38 
\bibitem[Janka(2012)]{2012ARNPS..62..407J} Janka, H.-T.\ 2012, Annual 
Review of Nuclear and Particle Science, 62, 407 
\bibitem[Katz et al.(2010)]{2010ApJ...716..781K} Katz, B., Budnik, R., 
\& Waxman, E.\ 2010, \apj, 716, 781 
\bibitem[Katz et al.(2012)]{2012ApJ...747..147K} Katz, B., Sapir, N., 
\& Waxman, E.\ 2012, \apj, 747, 147 
\bibitem[Klein 
\& Chevalier(1978)]{1978ApJ...223L.109K} Klein, R.~I., \& Chevalier, R.~A.\ 1978, \apjl, 223, L109 
\bibitem[Kotake et al.(2006)]{2006RPPh...69..971K} Kotake, K., Sato, K., 
\& Takahashi, K.\ 2006, Reports on Progress in Physics, 69, 971 
\bibitem[Krumholz et al.(2007)]{2007ApJ...667..626K} Krumholz, M.~R., 
Klein, R.~I., McKee, C.~F., \& Bolstad, J.\ 2007, \apj, 667, 626 
\bibitem[Levermore(1984)]{1984JQSRT..31..149L} Levermore, C.~D.\ 1984, 
\jqsrt, 31, 149
\bibitem[Matzner 
\& McKee(1999)]{1999ApJ...510..379M} Matzner, C.~D., \& McKee, C.~F.\ 1999, \apj, 510, 379
\bibitem[Matzner et al.(2013)]{2013ApJ...779...60M} Matzner, C.~D., Levin, 
Y., \& Ro, S.\ 2013, \apj, 779, 60  
\bibitem[Mazzali et al.(2006)]{2006Natur.442.1018M} Mazzali, P.~A., Deng, 
J., Nomoto, K., et al.\ 2006, \nat, 442, 1018 
\bibitem[Mignone 
\& Bodo(2005)]{2005MNRAS.364..126M} Mignone, A., \& Bodo, G.\ 2005, \mnras, 364, 126
\bibitem[Mihalas 
\& Mihalas(1984)]{1984oup..book.....M} Mihalas, D., \& Mihalas, B.~W.\ 1984, New York, Oxford University Press, 1984, 731 p., 
\bibitem[Moriya et al.(2011)]{2011MNRAS.415..199M} Moriya, T., Tominaga, 
N., Blinnikov, S.~I., Baklanov, P.~V., 
\& Sorokina, E.~I.\ 2011, \mnras, 415, 199 
\bibitem[Moriya et 
al.(2015)]{2015A&A...575L..10M} Moriya, T.~J., Sanyal, D., \& Langer, N.\ 2015, \aap, 575, L10 
\bibitem[Nakayama 
\& Shigeyama(2005)]{2005ApJ...627..310N} Nakayama, K., \& Shigeyama, T.\ 2005, \apj, 627, 31
\bibitem[Nakar 
\& Sari(2010)]{2010ApJ...725..904N} Nakar, E., \& Sari, R.\ 2010, \apj, 725, 904
\bibitem[Nakar 
\& Sari(2012)]{2012ApJ...747...88N} Nakar, E., \& Sari, R.\ 2012, \apj, 747, 88
\bibitem[Nomoto 
\& Hashimoto(1988)]{1988PhR...163...13N} Nomoto, K., \& Hashimoto, M.\ 1988, \physrep, 163, 13 
\bibitem[Ofek et al.(2010)]{2010ApJ...724.1396O} Ofek, E.~O., Rabinak, I., 
Neill, J.~D., et al.\ 2010, \apj, 724, 1396 
\bibitem[Ofek et al.(2013)]{2013Natur.494...65O} Ofek, E.~O., Sullivan, M., 
Cenko, S.~B., et al.\ 2013, \nat, 494, 65 
\bibitem[Ofek et al.(2014)]{2014ApJ...789..104O} Ofek, E.~O., Sullivan, M., 
Shaviv, N.~J., et al.\ 2014, \apj, 789, 104
\bibitem[Ohtani et al.(2013)]{2013ApJ...777..113O} Ohtani, Y., Suzuki, A., 
\& Shigeyama, T.\ 2013, \apj, 777, 113 
\bibitem[Pian et al.(2006)]{2006Natur.442.1011P} Pian, E., Mazzali, P.~A., 
Masetti, N., et al.\ 2006, \nat, 442, 1011 
\bibitem[Piro et al.(2010)]{2010ApJ...708..598P} Piro, A.~L., Chang, P., 
\& Weinberg, N.~N.\ 2010, \apj, 708, 598
\bibitem[Rabinak 
\& Waxman(2011)]{2011ApJ...728...63R} Rabinak, I., \& Waxman, E.\ 2011, \apj, 728, 63 
\bibitem[Roedig et al.(2012)]{2012MNRAS.426.1613R} Roedig, C., Zanotti, O., 
\& Alic, D.\ 2012, \mnras, 426, 1613 
\bibitem[Rybicki 
\& Lightman(1979)]{1979rpa..book.....R} Rybicki, G.~B., \& Lightman, A.~P.\ 1979, New York, Wiley-Interscience, 1979.~393 p., 
\bibitem[S{\c a}dowski et al.(2013)]{2013MNRAS.429.3533S} S{\c a}dowski, 
A., Narayan, R., Tchekhovskoy, A., \& Zhu, Y.\ 2013, \mnras, 429, 3533 
\bibitem[Sagiv et al.(2014)]{2014AJ....147...79S} Sagiv, I., Gal-Yam, A., 
Ofek, E.~O., et al.\ 2014, \aj, 147, 79 
\bibitem[Saio et al.(1988)]{1988Natur.334..508S} Saio, H., Nomoto, K., 
\& Kato, M.\ 1988, \nat, 334, 508 
\bibitem[Sakurai(1960)]{S60} Sakurai, A. 1960, Commun. Pure Appl. Math., 13, 353
\bibitem[Salbi et al.(2014)]{2014ApJ...790...71S} Salbi, P., Matzner, 
C.~D., Ro, S., \& Levin, Y.\ 2014, \apj, 790, 71
\bibitem[Sapir et al.(2011)]{2011ApJ...742...36S} Sapir, N., Katz, B., 
\& Waxman, E.\ 2011, \apj, 742, 36 
\bibitem[Sapir et al.(2013)]{2013ApJ...774...79S} Sapir, N., Katz, B., 
\& Waxman, E.\ 2013, \apj, 774, 79 
\bibitem[Sapir 
\& Halbertal(2014)]{2014ApJ...796..145S} Sapir, N., \& Halbertal, D.\ 2014, \apj, 796, 145
\bibitem[Schawinski et al.(2008)]{2008Sci...321..223S} Schawinski, K., 
Justham, S., Wolf, C., et al.\ 2008, Science, 321, 223 
\bibitem[Shigeyama et 
al.(1988)]{1988A&A...196..141S} Shigeyama, T., Nomoto, K., \& Hashimoto, M.\ 1988, \aap, 196, 141
\bibitem[Shigeyama 
\& Nomoto(1990)]{1990ApJ...360..242S} Shigeyama, T., \& Nomoto, K.\ 1990, \apj, 360, 242 
\bibitem[Soderberg et al.(2006)]{2006Natur.442.1014S} Soderberg, A.~M., 
Kulkarni, S.~R., Nakar, E., et al.\ 2006, \nat, 442, 1014 
\bibitem[Soderberg et al.(2008)]{2008Natur.453..469S} Soderberg, A.~M., 
Berger, E., Page, K.~L., et al.\ 2008, \nat, 453, 469 
\bibitem[Starling et al.(2011)]{2011MNRAS.411.2792S} Starling, R.~L.~C., 
Wiersema, K., Levan, A.~J., et al.\ 2011, \mnras, 411, 2792 
\bibitem[Stone et al.(1992)]{1992ApJS...80..819S} Stone, J.~M., Mihalas, 
D., \& Norman, M.~L.\ 1992, \apjs, 80, 819 
\bibitem[Suzuki 
\& Shigeyama(2010)]{2010ApJ...719..881S} Suzuki, A., \& Shigeyama, T.\ 2010, \apj, 719, 881
\bibitem[Suzuki 
\& Shigeyama(2010)]{2010ApJ...717L.154S} Suzuki, A., \& Shigeyama, T.\ 2010, \apjl, 717, L154
\bibitem[Svirski et al.(2012)]{2012ApJ...759..108S} Svirski, G., Nakar, E., 
\& Sari, R.\ 2012, \apj, 759, 108
\bibitem[Svirski \& Nakar(2014)]{2014ApJ...788..113S} Svirski, G., \& Nakar, E.\ 2014, \apj, 788, 113
\bibitem[Takahashi et al.(2013)]{2013ApJ...764..122T} Takahashi, H.~R., 
Ohsuga, K., Sekiguchi, Y., Inoue, T., \& Tomida, K.\ 2013, \apj, 764, 122
\bibitem[Takahashi 
\& Ohsuga(2013)]{2013ApJ...772..127T} Takahashi, H.~R., \& Ohsuga, K.\ 2013, \apj, 772, 127
\bibitem[Tan et al.(2001)]{2001ApJ...551..946T} Tan, J.~C., Matzner, C.~D., 
\& McKee, C.~F.\ 2001, \apj, 551, 946 
\bibitem[Tolstov et al.(2013)]{2013MNRAS.429.3181T} Tolstov, A.~G., 
Blinnikov, S.~I., \& Nadyozhin, D.~K.\ 2013, \mnras, 429, 318
\bibitem[Tominaga et al.(2009)]{2009ApJ...705L..10T} Tominaga, N., 
Blinnikov, S., Baklanov, P., et al.\ 2009, \apjl, 705, L10
\bibitem[Tominaga et al.(2011)]{2011ApJS..193...20T} Tominaga, N., 
Morokuma, T., Blinnikov, S.~I., et al.\ 2011, \apjs, 193, 20
\bibitem[Turner 
\& Stone(2001)]{2001ApJS..135...95T} Turner, N.~J., \& Stone, J.~M.\ 2001, \apjs, 135, 95 
\bibitem[van Leer(1977)]{1977JCoPh..23..276V} van Leer, B.\ 1977, Journal 
of Computational Physics, 23, 276
\bibitem[Wang et al.(2007)]{2007ApJ...664.1026W} Wang, X.-Y., Li, Z., 
Waxman, E., \& M{\'e}sz{\'a}ros, P.\ 2007, \apj, 664, 1026 
\bibitem[Wang 
\& Wheeler(2008)]{2008ARA&A..46..433W} Wang, L., \& Wheeler, J.~C.\ 2008, \araa, 46, 433 
\bibitem[Waxman et al.(2007)]{2007ApJ...667..351W} Waxman, E., 
M{\'e}sz{\'a}ros, P., \& Campana, S.\ 2007, \apj, 667, 351 
\bibitem[Weaver(1976)]{1976ApJS...32..233W} Weaver, T.~A.\ 1976, \apjs, 32, 
233 
\bibitem[Zhang et al.(2011)]{2011ApJS..196...20Z} Zhang, W., Howell, L., 
Almgren, A., Burrows, A., \& Bell, J.\ 2011, \apjs, 196, 20
\end{thebibliography}
\end{document}